\documentclass[superscriptaddress,
reprint,longbibliography,
 amsmath,amssymb,
aps,
pra,
]{revtex4-1}

\usepackage{simpler-wick}
\usepackage{graphicx}
\usepackage{dcolumn}
\usepackage{bm}
\usepackage{subfig}
\usetikzlibrary{decorations.markings}
\usetikzlibrary{arrows.meta,positioning,calc,fit,matrix,decorations.pathmorphing,decorations.markings,shapes.geometric}
\usetikzlibrary{shapes.misc}

\usepackage{tikz}
\usepackage{graphicx}
\usepackage{subfig}
\usetikzlibrary{arrows.meta}

\usepackage{subcaption}
\usepackage{silence}

\usepackage[colorlinks=true,linkcolor=blue,citecolor=blue,urlcolor=blue]{hyperref}
\usepackage{braket}
\usepackage{bbm}

\usepackage{amsthm}

\usepackage{feynmf}

\begin{document}


\title{Irreversibility, equilibrium and measurement in quantum mechanics}

\author{Christopher J. N. Coveney}
\email{christopher.coveney@physics.ox.ac.uk}
\affiliation{ Department of Physics, University of Oxford, Oxford, OX1 3PJ, United Kingdom}
\author{Peter V. Coveney}%
\email{p.v.coveney@ucl.ac.uk}
\affiliation{Centre for Computational Science, University College London, London, WC1H 0AJ, United Kingdom}
\affiliation{Advanced Research Computing Centre, University College London, London, WC1H 0AJ, United Kingdom}

\date{\today}

\begin{abstract}
    Quantum mechanics is widely recognised as being incomplete. It is not consistent with the second law of thermodynamics and does not provide a scientifically credible physical account of the measurement process, the means by which coherence is broken and classically observable states are recorded. This has led to many \emph{ad hoc} assumptions being used to account for various properties of quantum systems, among which is the coherence time of quantum devices that determines their ability to perform computations. Here, we show that all these properties can be accommodated naturally and consistently in the context of mixing quantum systems which exhibit continuous spectra, as arises in the thermodynamic limit in quantum statistical mechanics and quantum gravity. In particular, for isolated systems we show that the time-reversal symmetry associated with unitary time evolution of the quantum state gives rise to time-symmetry breaking and a semi-group evolution which attains thermodynamic equilibrium at long times. Moreover, the emergence of this non-unitary time-asymmetry leads to microcanonical equilibrium states in which all quantum coherence is lost and is accompanied by the transformation of pure states into mixtures, leading in turn to an increase in entropy.  Inclusion of a macroscopic measurement apparatus shows how the outcome of a measurement corresponds to the von Neumann projection postulate, arising with probabilities in conformance with the Born rule. The mathematical structure of the theory which applies to quantum systems with continuous spectra is closely analogous to the classical ergodic theory of dynamical systems and the conditions under which they attain equilibrium states. 
\end{abstract}


\maketitle

\begin{fmffile}{diagram}

\section{Introduction}

Quantum mechanics as it is known today is an incomplete theory in a number of respects. The time-reversal symmetry of its equations of motion prevents it from exhibiting the irreversibility dictated by the second law of thermodynamics, and this problem is aligned with that which arises in the measurement problem, as the decoherence leading to wave function collapse is associated with an increase in entropy~\cite{von2018mathematical}. The lack of an internally consistent solution is the reason why there are so many diverse interpretations and attempted explanations of the measurement problem, all of which are imposed on the theory from the outside with varying degrees of subjectivity~\cite{heisenberg2025physics,bohr1928quantum,schrodinger1935present,everett1957hugh,von2018mathematical,dirac1981principles}. Here, we show that the solution resides within the theory itself: by studying the statistical mechanics of large quantum systems, we demonstrate that — for quantum systems with absolutely continuous spectra — the time symmetry is broken and the temporal evolution leads to the equilibrium microcanonical distribution. The irreversibility causes coherence to be lost, pure states become mixtures, wave functions collapse, and the entropy increases to its maximum value in the equilibrium state, as demanded by the second law of thermodynamics.  

We show that all these consequences flow from a representation of quantum mechanics in terms of the algebraic quantum state, which is governed by the Liouville-von Neumann (LvN) equation.  The approach is closely analogous to the classical ergodic theory of dynamical systems, where the analysis is framed in terms of the Liouville equation which describes the time evolution of the entire distribution function. Both approaches make use of complex variable theory to solve the Liouville or LvN equation in terms of an inverse Laplace transform in which the key object whose analytic properties are of interest is the resolvent $(z-\mathcal{L})^{-1}$, where $\mathcal{L}$ is the Liouville operator or Liouvillian. In both the classical and the quantum theory, the requirement for an approach to equilibrium is that the spectrum of $\mathcal{L}$ must be absolutely continuous, a property synonymous with mixing in classical mechanics, which in turn requires the classical dynamics to be chaotic~\cite{coveney2025molecular}. In a Newtonian, trajectory-based, description of classical mechanics, all isolated ergodic systems undergo Poincaré recurrence~\cite{poincare1890probleme}, whereby the system eventually (on an astronomical timescale) returns arbitrarily close to its initial state; in the probabilistic description, however, for a mixing system the distribution function reaches a time-independent state, namely equilibrium~\cite{gaspard2005chaos}. In the quantum-mechanical case, the analogue of the trajectory is a wave function, while the probability distribution function corresponds to the density matrix. Since for all finite quantum systems the spectrum of $\mathcal{L}$ is discrete, they undergo a strong form of Poincaré recurrence as not only the wave function but also the density matrix is almost periodic. It is only in the infinite system limit where quantum systems exhibit continuous spectra, and this is the case when we are studying systems in the thermodynamic limit (where both the number of particles $N$ and the volume of the system $V$ become infinite in such a way that the density, $\frac{N}{V}$, remains constant).  In this limit, the properties of quantum mechanical systems have many parallels with those classical dynamical systems which exhibit equilibrium states. 

We start by looking at the structure of the LvN equation in the case where the spectrum of the Liouvillian is absolutely continuous. In this situation, which has been rarely studied, the analytic properties of the resolvent are much more complicated than for finite-sized systems and the form of analytical continuation necessary to analyse the dynamics is substantially more involved. We develop the necessary mathematical foundations on which to study such dissipative isolated quantum systems, and go on to reveal the irreversible nature of the time evolution that emerges by time reversal symmetry breaking. This in turn leads us to derive equations of motion that evolve towards microcanonical equilibrium states. In the case of quantum measurement, we find that the approach to equilibrium produces mixed states with probabilities given by the Born rule, providing a rigorous explanation of von Neumann’s projection postulate for the collapse of wave functions.

\section{Time Evolution of observables in quantum mechanics}

In quantum mechanics, the time-evolution of observables is carried out in the Heisenberg picture:
\begin{gather}
    \begin{split}~\label{eq:LvN_heisenberg}
        \frac{\partial}{\partial t} A_t = i\mathcal{L}A_t \ ,
    \end{split}
\end{gather}
where $\mathcal{L}$ is the Liouville operator that generates the dynamics of the system. Here we have assumed that the Hamiltonian and observables do not carry an explicit time-dependence. Within a Hilbert space representation, we have $\mathcal{L}\equiv [H,\cdot]$, where $H$ is the Hamiltonian of the system. In Eq.~\ref{eq:LvN_heisenberg}, the Liouvillian acts on the abstract Banach space such that $\mathcal{L}:\mathcal{A}\to\mathcal{A}$, where $\mathcal{A}$ represents the Banach algebra of observables (see Appendix~\ref{app:aqm}). Algebraically, the time-dependent average of an observable obeys the relationship: $\omega(A_t)=\omega_{t}(A)$, where $\omega$ is the quantum state given by the functional $\omega : \mathcal{A}\to\mathbbm{C}$~\cite{davies1976quantum,fewster2020algebraic}. The modern algebraic formulation of quantum theory has found extensive application in quantum gravity and quantum statistical mechanics research~\cite{witten2018aps,witten2022does,furuya2023information}. In particular, Witten explicitly showed that the thermodynamic limit requires the algebraic quantum theory to employ Type III von Neumann algebras which do not possess trace class representations~\cite{witten2018aps,witten2022does}.
Using the Schrödinger picture, we can write the conjugate evolution of the time-dependent state functional explicitly in terms of the abstract Liouville-von Neumann equation
\begin{gather}
    \begin{split}~\label{eq:LvN}
        \frac{\partial}{\partial t} \omega_{t} = -i \mathcal{L}\omega_{t} \ , 
    \end{split}
\end{gather}
where now the Liouvillian generator acts on the dual Banach space $\mathcal{A^*}$ as $\mathcal{L}:\mathcal{A^*}\to\mathcal{A^*}$. 
This equation encodes the change in the quantum state of the system, $\omega_{t}$. Clearly, Eqs~\ref{eq:LvN_heisenberg} and~\ref{eq:LvN} require $[-\mathcal{L}\omega_{t}](A)=\omega_{0}(\mathcal{L}A_t)$. Taking the Laplace transform of the LvN equation by defining $\tilde{\omega}_{s} = \int^{\infty}_{0} dt\  e^{-st}\omega_{t}$, we have
\begin{gather}
    \begin{split}
        \tilde{\omega}_{s}&= (s\mathbbm{1}+i\mathcal{L})^{-1}\omega_0 , 
    \end{split}
\end{gather}
where $\omega_{0}$ is the initial quantum state at $t=0$. 
We can then invert this expression using the inverse Laplace transform (Bromwich integral) to give the general solution $(t>0)$:
\begin{gather}
    \begin{split}
        \omega_{t} 
        &= \int^{\gamma+i\infty}_{\gamma-i\infty} \frac{ds}{2\pi i}\  e^{st}(s\mathbbm{1}+i\mathcal{L})^{-1}\omega_0 \ .
    \end{split}
\end{gather}
Here the contour of integration is a straight line parallel to the imaginary axis with all singularities to the left of the resolvent $(s+i\mathcal{L})^{-1}$. 
Performing a Wick rotation $z=is$, we can re-write the expression as 
\begin{gather}
    \begin{split}
        \omega_{t} &=i\int^{\infty+i\gamma}_{-\infty+i\gamma} \frac{dz}{2\pi }\  e^{-izt}(z\mathbbm{1}-\mathcal{L})^{-1}\omega_0 \ , 
    \end{split}
\end{gather}
where now the singularities of the resolvent are located on the real axis. Using this result, we can immediately write the time-dependent expectation value of an observable $A$ in the quantum state $\omega_{t}$ as 
\begin{gather}
    \begin{split}~\label{eq:time_av}
        \omega_{t}(A) &=i\int^{\infty+i\gamma}_{-\infty+i\gamma} \frac{dz}{2\pi}\  e^{-izt} G_{A}(z)\ ,
    \end{split}
\end{gather}
where we have introduced the Green's function $G_{A}(z) = [(z\mathbbm{1}-\mathcal{L})^{-1}\omega_0](A)=\omega_{0}((z\mathbbm{1}+\mathcal{L})^{-1}A)$. 
Collecting both time directions into the definition of the inverse, we can write 
\begin{gather}
    \begin{split}~\label{eq:time_av_cont}
        \omega_{t}(A) &=\oint_{\mathcal{C}} \frac{dz}{2\pi i}\  e^{-izt} G_{A}(z)\ ,
    \end{split}
\end{gather}
where the contour $\mathcal{C}$ chosen depends on whether $t\gtrless 0$. 

From Eqs~\ref{eq:time_av} and~\ref{eq:time_av_cont}, it is clear that the time evolution of observables in quantum theory is deeply connected to the structure of the resolvent of the Liouvillian: $R(z) = (z\mathbbm{1}-\mathcal{L})^{-1}$. As we will demonstrate, the emergence of irreversibility, the arrow of time, the approach to equilibrium as well as the quantum measurement process fundamentally depend on the analytical structure of this resolvent.

\section{The thermodynamic limit and ensembles in statistical mechanics}

The thermodynamic limit is one of, if not the most, important concepts in statistical mechanics. Consider an isolated system consisting of $N$ interacting particles in a volume $V$. The thermodynamic limit is defined as the limit where we take $N,V\to\infty$, requiring the density to remain constant, $\rho = \frac{N}{V}= C$. This limit is essential in order to mathematically define phase transitions and spontaneous symmetry breaking in physics. For example, nonanalyticities in the free energy are only possible after taking the thermodynamic limit. The phenomena of phase transitions and spontaneous symmetry breaking are objective and experimentally observable. Therefore, the thermodynamic limit is essential in order to describe phenomena exhibited by macroscopically large numbers of interacting particles. In the thermodynamic limit, the spectrum of the Hamiltonian and Liouvillian becomes dense and continuous. In many cases, the spectrum of the Liouvillian becomes absolutely continuous. It is important to note that quantum field theory and scattering theory also both depend heavily on the existence of continuous spectra~\cite{Quantum,dzyaloshinski1975methods,mahan2000many,peskin2018introduction}. Without the continuum of energy eigenstates, it would be impossible to provide a description of scattering rates and subsequent transitions between states in quantum mechanics~\cite{coveney2026thesis}. As we will demonstrate, it is the absolutely continuous nature of the spectrum of the Liouvillian that gives rise to irreversible decay and the approach to equilibrium in quantum mechanics. The broken time-symmetry is related to the non-commutativity of the two limits:
\begin{gather}
    \begin{split}~\label{eq:broken_symm}
       \lim_{t\to\infty}\lim_{\substack{N,V\to\infty\\\frac{N}{V}\to C}} f_{N,V}(t) \neq \lim_{\substack{N,V\to\infty\\\frac{N}{V}\to C}}\lim_{t\to\infty} f_{N,V}(t) \ , 
    \end{split}
\end{gather}
where $f_{N,V}(t)$ is a time-dependent function. The left hand side of Eq.~\ref{eq:broken_symm} requires the thermodynamic limit to be taken first before the asymptotic time evolution, whereas the right hand side reverses these limits. The limit is singular because the order in which these two limits are taken, $\lim_{t\to\infty}\lim_{N,V\to\infty}$, gives totally different results: it is the first limit which is the physically correct one as it describes the asymptotic time evolution of a macroscopic number of interacting particles. This is analogous to the non-commuting limits that give rise to spontaneous symmetry breaking in a magnetic material: $\lim_{h\to 0}\lim_{N,V\to\infty}Z\neq\lim_{N,V\to\infty}\lim_{h\to 0}Z $, where $h$ represents a weak external field and $Z$ is the partition function~\cite{anderson1958absence,anderson1961localized,anderson1963plasmons,anderson1972more,altland2010condensed}. Even though the microscopic dynamics is reversible, we will demonstrate that the thermodynamic limit, required in order to correctly describe macroscopically large numbers of interacting particles, breaks this time-symmetry. The thermodynamic limit provides the only mathematical structure capable of reflecting the objective reality of emergence in many-particle physics. Without it, finite equations would insist that ice never truly melts and water never truly boils, in direct contradiction with experimental observation. In particular, Anderson's Nobel prize-winning work on the electronic structure of magnetic and disordered systems is fundamentally dependent on the thermodynamic limit in order to explain spontaneous symmetry breaking in antiferromagnetic materials~\cite{anderson1958absence,anderson1961localized}. Furthermore, his analysis of spontaneous symmetry breaking in superconductors demonstrated that photons can acquire mass in the interior of the superconductor~\cite{anderson1963plasmons,anderson1972more}. Anderson's contribution laid the groundwork for the development of the famous Higgs mechanism in high energy physics~\cite{higgs1964broken,englert1964broken}.

As stated in the introduction, finite quantum systems display a strong form of Poincaré recurrence~\cite{poincare1890probleme,bocchieri1957quantum}. This is due to the fact that their spectra are fundamentally discrete. For a finite system containing $N$ particles, the Poincaré recurrence time, $\mathcal{T}_{\text{PR}}(N)$, is the time taken for the state of a finite quantum system to return to its initial configuration. In general, $\mathcal{T}_{\text{PR}}(N)\sim e^{N}$ scales at least exponentially with the size of the system. For typical  quantum systems containing macroscopic numbers of particles ($N\sim10^{23}$), the Poincaré recurrence time vastly exceeds the age of the universe. In particular, as the size of the system increases, $\mathcal{T}_{\text{PR}}(N)\to \infty$. As a quantum system increases in size, the spacing between the adjacent energy levels typically decreases exponentially: $\Delta E(N) \sim e^{-N}$. This discrete level structure can only be resolved over time scales that are long in comparison to $\frac{\hbar}{\Delta E(N)}$. As $\Delta E(N)$ becomes vanishingly small for a macroscopic system, an observation taken over a typical time scale $\tau_{\text{obs}}$ generally satisfies the inequality $\Delta E(N) \ll \frac{\hbar}{\tau_{\text{obs}}}$. This means that an experiment can only detect the level density averaged over the experimentally accessible energy interval: $\frac{\hbar}{\tau_{\text{obs}}}$~\cite{Quantum}. As a result, the spectral measure of a typical macroscopic quantum system is operationally indistinguishable from a continuum. This is captured by the fact that the  corresponding sequence of spectral measures converges to a continuous measure in the thermodynamic limit. At the same time, as the size of the system increases, $\mathcal{T}_{\text{PR}}(N)$ diverges thereby removing finite-size spectral resolution and Poincaré recurrence from every fixed observational time window~\cite{maldacena2003eternal,barbon2003very,festuccia2007arrow,furuya2023information}. Therefore, the singular thermodynamic limit allows for the extrapolation of the asymptotic behaviour exhibited by macroscopically large numbers of interacting particles and provides the correct physically motivated limit to describe macroscopic quantum systems. 

In addition to the thermodynamic limit, statistical mechanics is rooted in ensemble theory~\cite{tolman1979principles,ruelle2004thermodynamic}. The most fundamental ensemble in statistical mechanics is the microcanonical ensemble, which corresponds to an \emph{isolated system} at fixed number of particles, total volume and energy $(N,V,E)$~\cite{tolman1979principles}. The ensemble assumes that only the total energy is conserved by the dynamics. In terms of the density matrix, it is given by the \emph{distribution}
\begin{gather}
    \begin{split}
        \rho^{\text{mce}}_{\text{eq}}(E) =\frac{\delta(H-E)}{\Omega(E)} \ ,
    \end{split}
\end{gather}
where $\Omega(E) = \text{Tr}\{\delta(H-E)\}$ and $H$ is the Hamiltonian of the total closed system. In the context of classical mechanics, $\rho^{\text{mce}}_{\text{eq}}(E)$, is often referred to as the \emph{invariant measure} as it is stationary under the time evolution of the system. In particular, dissipative systems emphasise the importance of singular invariant measures for which $\rho_{\text{eq}}$ is a distribution as is the case for an attracting stationary point, for example $\rho_{\text{eq}}(X)=\delta(X-X_s)$~\cite{gaspard2005chaos}. The microcanonical ensemble ensures that the system is restricted to lie on a specific energy shell corresponding to the various microstates compatible with this energy constraint. Historically, the microcanonical ensemble was initially posited by Boltzmann~\cite{boltzmann1877beziehung} and Maxwell~\cite{maxwell1879boltzmann}, being further developed by Einstein~\cite{einstein1902kinetisches} and Gibbs~\cite{gibbs1902elementary}.  

Alongside the microcanonical ensemble, Gibbs also introduced the concept of the canonical ensemble by imagining that the full system of fixed $(N,V,E)$ is embedded within an even larger system (consisting of the system and the surroundings). In this case, the system under consideration is referred to as a \emph{closed system}. In order that the system and its surroundings are in equilibrium through only the transfer of energy requires the temperature of the system and surroundings to be constant (along with the volume and number of particles). As a result, the canonical ensemble corresponds to a system in equilibrium with its surroundings through only the exchange of energy such that the closed system has a constant number of particles, volume and temperature $(N,V,T)$. This situation is described by the Boltzmann distribution 
\begin{gather}
    \begin{split}
        \rho^{\text{ce}}_{\text{eq}}(\beta) = \frac{e^{-\beta H}}{Z_\text{ce}(\beta)}\ ,
    \end{split}
\end{gather}
where $Z_\text{ce}(\beta)=\text{Tr}\{e^{-\beta H}\}$ is the canonical partition function. The inverse temperature, $\beta=\frac{1}{k_BT}$, is determined by the entropy change of the system as its energy fluctuates due to interaction with the surroundings. 

We can extend the concept of the canonical ensemble by allowing the particles as well as energy to be exchanged with the surroundings. This defines the grand canonical ensemble and the system it describes is referred to as an \emph{open system}. In this case, equilibrium between the open system and the surroundings is obtained when both the temperature, $T$, and the chemical potential, $\mu$, are constant. Analogously to the temperature, the chemical potential is determined by the entropy change of the system as its particle number fluctuates due to exchange with the surroundings (keeping the volume and internal energy of the system constant). As a result, the grand canonical ensemble corresponds to a system in equilibrium with its surroundings through the exchange of energy and particles such that the open system has a constant chemical potential, volume and temperature $(\mu,V,T)$. The generalized Boltzmann distribution corresponding to an open system at equilibrium is given by 
\begin{gather}
    \begin{split}
        \rho^{\text{gce}}_{\text{eq}}(\beta,\mu) = \frac{e^{-\beta (H-\mu N)}}{Z_\text{gce}(\beta,\mu)}\ ,
    \end{split}
\end{gather}
where $Z_\text{gce}(\beta,\mu)=\text{Tr}\{e^{-\beta (H-\mu N)}\}$ is the grand canonical partition function. As we shall discuss in Section~\ref{sec:measurement}, quantum decoherence theory is fundamentally dependent on the concept of an open system introduced by the grand canonical ensemble.

In the thermodynamic limit and at equilibrium, it was rigorously shown that the microcanonical, canonical and grand canonical ensembles give the same result for the expectation value of any local observable due to the fact that fluctuations decay as $\mathcal{O}(\frac{1}{\sqrt{N}})$~\cite{ruelle1969statistical,lanford1969observables,ruelle2004thermodynamic}. The concept of the thermodynamic limit is crucial in order to establish the equivalence between the different ensembles employed in statistical mechanics. However, we note that the inverse square root law for fluctuations is not always true for certain `complex systems' such as those near continuous phase transitions~\cite{wilson1971renormalization,wilson1971renormalization2}.

\section{Spectral properties of the Liouvillian and topology of the resolvent}

In quantum mechanics, the commutator structure of the Liouvillian means that the spectrum of the Liouvillian $\sigma(\mathcal{L})$ is closely connected to the spectrum of the Hamiltonian $\sigma(H)$. In particular, the Hilbert space spectrum of the Liouvillian is given by~\cite{antoniou2003irreversibility}
\begin{gather}
    \begin{split}~\label{eq:spec_liouville}
        \sigma(\mathcal{L}) = \{\lambda-\lambda' : \lambda, \lambda' \in \sigma(H)\} \ .
    \end{split}
\end{gather}
Clearly $\sigma(\mathcal{L})$ consists of energy differences in the spectrum of the Hamiltonian and is therefore symmetric. For example, if the spectrum of the Hamiltonian consists of the positive real line $\mathbbm{R}^+ = [0,\infty)$, the Liouvillian spectrum is the entire real line, $\mathbbm{R}$. The continuous spectrum of the Liouvillian does not necessarily require the thermodynamic limit and commonly occurs in quantum systems such as quantum fields and throughout scattering theory~\cite{reed1979iii}. It is also possible that the spectrum of the Liouvillian can give rise to richer structures than that displayed by the Hamiltonian~\cite{antoniou2003irreversibility}. It is clear that the nature of the spectrum of the Liouvillian controls the topology of the resolvent $R(z)$ which will have a significant impact on the nature of the time-evolution of observables through Eq.~\ref{eq:time_av}. In this paper, we will discuss three main cases relevant to non-integrable quantum systems:
\begin{enumerate}
    \item \textbf{Periodic--} Pure point spectrum: $\sigma(\mathcal{L})=\sigma_{\text{p}}(\mathcal{L})$. Typically this corresponds to the spectrum of a finite quantum system. 
    \item \textbf{Quasi-periodic--} Point spectrum and continua that begin above certain energy thresholds: $\sigma(\mathcal{L})= \sigma_{\text{p}}(\mathcal{L})+\sigma_{\text{ac}}(\mathcal{L})$. 
    \item \textbf{Mixing--} Absolutely continuous spectrum: $\sigma(\mathcal{L})=\sigma_{\text{ac}}(\mathcal{L})$. 
\end{enumerate}
There are also cases where the spectrum could contain singular continuous contributions~\cite{antoniou2003irreversibility}. However, these contributions are not relevant to the present work. The above three different cases result in fundamentally different analytical structures of the resolvent and therefore directly influence the dynamics of observables, states and time-symmetry in quantum mechanics. 

We wish to point out that in the theory of dynamical systems, mixing can be equivalently formulated through the condition~\cite{gaspard2005chaos,coveney2025molecular,bach2000return,festuccia2007arrow}
\begin{gather}
    \begin{split}~\label{eq:mixing_def}
        \lim_{t\to\infty} \omega_{\text{i}}(A_tB) = \omega_{\text{i}}(A)\omega_{\text{i}}(B) \ , 
    \end{split}
\end{gather}
where $\omega_{\text{i}}$ is referred to as the invariant measure that is preserved by the dynamics $(\frac{\partial}{\partial t}\omega_{\text{i}}=0)$. Importantly, $\omega_{\text{i}}$ must satisfy the conditions required in order to be an algebraic state (see Appendix~\ref{app:aqm}). Eq.~\ref{eq:mixing_def} requires the spectrum of the non-integrable Liouvillian to be absolutely continuous~\cite{maldacena2003eternal,festuccia2007arrow,furuya2023information}. Quantum field theory methods are commonly based around the analysis of dynamical correlation functions of the form 
\begin{gather}
    \begin{split}
        G_{AB}(t) = \omega_{\text{i}}(A_tB) - \omega_{\text{i}}(A_t)\omega_{\text{i}}(B) \ , 
    \end{split}
\end{gather}
where $A$ and $B$ are observables. Dynamical correlation functions of this form are almost always assumed to satisfy the asymptotic condition, $\lim_{t\to\infty} G_{AB}(t)=0$, which means that we can take their Fourier transform to analyse the properties of the system in terms of the spectral function: $G_{AB}(\omega)$. Using Eq.~\ref{eq:mixing_def}, it is straightforward to show that mixing dynamical systems give rise to dynamical correlation functions that satisfy: $\lim_{t\to\infty} G_{AB}(t)=0$. Therefore, the mixing condition is actually implicitly assumed throughout quantum field theory, condensed matter physics as well as scattering theory. The assumption is made because it is exceptionally difficult to rigorously demonstrate the mixing property for most LvN operators as is also true in the classical case. However, in statistical mechanics the mixing condition is a necessary and sufficient condition for a dynamical system to exhibit an equilibrium state which is in turn a property of all thermodynamic systems. Therefore, we shall follow the underlying assumptions commonly employed in quantum field theory and condensed matter physics by assuming that the mixing property holds in all macroscopic non-integrable quantum systems of interest unless it can be proven otherwise~\cite{coveney2025molecular}. Unlike for classical dynamical systems, the mixing condition in quantum mechanics plainly only applies in the thermodynamic limit (when systems exhibit continuous spectra). The relationship between the mixing condition and the continuous spectrum of the generator of the dynamics is discussed extensively in the quantum gravity literature involving the thermodynamic properties of black holes~\cite{maldacena2003eternal,barbon2003very,festuccia2007arrow,furuya2023information}.

Of particular relevance in the theory of dynamical systems is the unitary Markovian time-evolution group 
\begin{gather}
    \begin{split}
        \hat{U}(t) = e^{-i\mathcal{L}t} \ , 
    \end{split}
\end{gather}
which is invertible for all $t\in\mathbbm{R}$.
Using the Gelfand–Naimark–Segal theorem~\cite{gelfand1943imbedding,segal1947irreducible}
we can express $\hat{U}(t)$ as acting on the Liouville-Hilbert space. Commonly, $\hat{U}(t)$ is referred to as the Frobenius-Perron operator. In terms of the algebraic formulation of quantum mechanics, $\hat{U}(t)$ acts on the dual Banach space of states. The unitary group is directly related to the resolvent of the Liouvillian through the inverse Laplace transformation as 
\begin{gather}
    \begin{split}
        \hat{U}(t) = i\int^{\infty+i\gamma}_{-\infty+i\gamma}\frac{dz}{2\pi}\ e^{-izt}(z\mathbbm{1}-\mathcal{L})^{-1} \ . 
    \end{split}
\end{gather}
Therefore, its structure is strongly influenced by the topology of the resolvent. For periodic systems, the unitary group is completely recovered for all $t\in \mathbbm{R}$ from the inverse Laplace transform (see Appendix~\ref{app:1})
\begin{gather}
    \begin{split}
        \oint_{\mathcal{C}}\frac{dz}{2\pi i}\ e^{-izt}(z\mathbbm{1}-\mathcal{L})^{-1} = e^{-i\mathcal{L}t} \hspace{2.5mm}\forall\hspace{1mm} t\in\mathbbm{R} \ .
        \end{split}
\end{gather}
This is because the simple pole analytic structure of the resolvent allows the inverse Laplace transform to be written in terms of a single closed contour for all $t$. Therefore, the time evolution of the system oscillates indefinitely and does not reach a stationary equilibrium state. This means that the asymptotic limit, $\lim_{t\to\infty}\omega_{t}(A)$, does not converge to a stationary state for periodic and quasi-periodic quantum systems. For periodic systems this is a consequence of a strong form of Poincaré recurrence, while the time oscillations of a quasi-periodic system never fully dissipate as the discrete point components of the spectrum continue to oscillate unitarily.

The relationship between a mixing system and the necessity of a continuous spectrum was first identified by Koopman and von Neumann~\cite{koopman1932dynamical}. The continuous spectrum of the Liouvillian leads to a fundamentally different structure of the unitary group for the case of mixing systems. This is because the resolvent of the non-integrable Liouvillian develops a discontinuity across the entire real axis:
\begin{gather}
    \begin{split}~\label{eq:discont}
       (\omega\mathbbm{1}-\mathcal{L}-i0^+)^{-1}-(\omega\mathbbm{1}-\mathcal{L}+i0^+)^{-1} = 2\pi i \Gamma(\omega) \ . 
    \end{split}
\end{gather} 
This discontinuity is exactly the spectral function: $\Gamma(\omega)=\delta(\omega-\mathcal{L})$. The notation $i0^{+}$ means evaluation of the resolvent an infinitesimal distance above the real axis.  The discontinuity means that the resolvent is not a single-valued function and so must contain a branch cut structure across the entire real axis. This results in a fundamentally different resolvent topology than that displayed for finite quantum systems with discrete point spectra. 
Importantly, the spectral function is entirely determined from the imaginary part of the retarded component through the relation:
\begin{gather}
    \begin{split}~\label{eq:dirac_rel}
        -\frac{1}{\pi}\text{Im}(\omega\mathbbm{1}-\mathcal{L}+i0^+)^{-1}=\Gamma(\omega) \ , 
    \end{split}
\end{gather}
with the imaginary part of the retarded component obeying the negative semi-definite property~\cite{stefanucci2013nonequilibrium,coveney2026thesis}. These relations are crucial in order for the theory to maintain causality. In particular, Eq.~\ref{eq:dirac_rel} underlies the Kramers-Kronig relations~\cite{kramers1925behavior,de1926theory}. The positive semi-definite property of Eq.~\ref{eq:dirac_rel} is a consequence of the Wiener-Khinchin theorem.

The presence of the branch cut directly for mixing systems leads to the strict separation of the inverse Laplace transform for time moving forward or backwards. As a result, the unitary group of the time evolution operator splits
into two semi-groups, corresponding to either strict forward (retarded) or backward (advanced) time evolution. This is the case because the topology of the resolvent prevents the inverse Laplace transform from being written in terms of a single closed contour for all $t$. Evaluation of the Bromwich integral for $t>0$ gives 
(see Appendix~\ref{app:spec_cont} and~\ref{app:continuation})
\begin{gather}
    \begin{split}
        i\int^{\infty}_{-\infty}\frac{d\omega}{2\pi}\ e^{-i\omega t}(\omega\mathbbm{1}-\mathcal{L}+i0^+)^{-1} &= \theta(t)e^{-i\mathcal{L}t}\\
        &=\hat{U}_{+}(t) \ .
    \end{split}
\end{gather}
This result is obtained by integration just above the branch cut discontinuity. 
For $t<0$, we have the corresponding expression 
\begin{gather}
    \begin{split}
        -i\int^{\infty}_{-\infty}\frac{d\omega}{2\pi}\ e^{-i\omega t}(\omega\mathbbm{1}-\mathcal{L}-i0^+)^{-1} &= \theta(-t)e^{-i\mathcal{L}t} \\
        &= \hat{U}_{-}(t) \ ,
    \end{split}
\end{gather}
where the integral is taken an infinitesimal distance beneath the branch cut. 
The emergence of the Heaviside step functions $\theta(\pm t)$ is enforced by the discontinuity of the resolvent across the entire real axis (see Eq.~\ref{eq:discont}). This is because the transforms involve operators, $(\omega\mathbbm{1}-\mathcal{L}\pm i0^{+})^{-1}$, which belong to operator-valued Hardy spaces and therefore only have support in the upper/lower half of the complex plane~\cite{duren1970theory,koosis1998introduction}. As a result, they only have time support for $t\gtrless0$, respectfully. It is important to note that it is the intrinsic topology of the resolvent for systems with continuous spectra that generates the causal step functions. This subtlety is commonly not appreciated in the literature on quantum field theory where the step functions are often introduced \emph{a priori} in the definition of projected resolvents (Green's function theory)~\cite{Quantum,dzyaloshinski1975methods,mahan2000many,altland2010condensed,stefanucci2013nonequilibrium}.

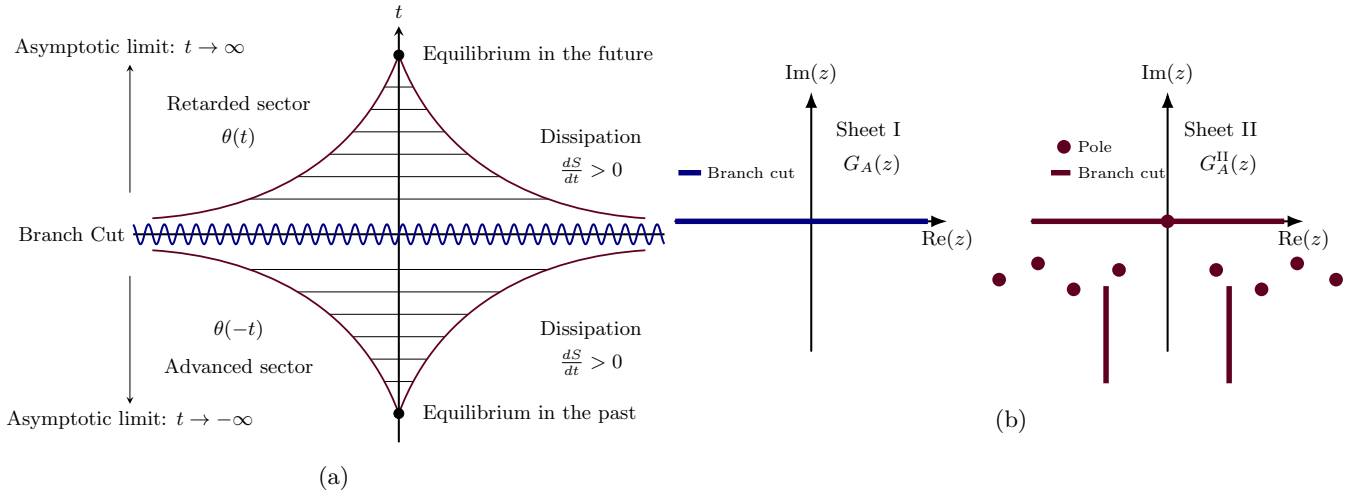
\begin{figure*}[t]
\centering

\begin{minipage}{0.5\textwidth}
\centering
\resizebox{\linewidth}{!}{%
\begin{tikzpicture}[>=stealth,scale=1.0]

\draw[->, line width=0.9pt] (0,-3.25) -- (0,3.25) node[above] {$t$};
\draw[-, line width=0.9pt] (4.15,0) -- (-4.15,0) node[left] {\hspace{1.5mm} Branch Cut};

\begin{scope}
\clip
(-3.85,0.25)
  .. controls (-2.25,0.35) and (-0.65,1.05) .. (0,2.8)
  .. controls (0.65,1.05) and (2.25,0.35) .. (3.85,0.25)
  --
(3.85,-0.25)
  .. controls (2.25,-0.35) and (0.65,-1.05) .. (0,-2.8)
  .. controls (-0.65,-1.05) and (-2.25,-0.35) .. (-3.85,-0.25)
  -- cycle;

\foreach \y in {-2.30,-1.95,-1.60,-1.25,-0.90,-0.55,
                 0.55,0.90,1.25,1.60,1.95,2.30}
{
  \draw[thin] (-4.0,\y) -- (4.0,\y);
}
\end{scope}

\draw[thick,draw=purple!50!black]
  (-3.85,0.25)
  .. controls (-2.25,0.35) and (-0.65,1.05) .. (0,2.8)
  .. controls (0.65,1.05) and (2.25,0.35) .. (3.85,0.25);

\draw[thick,draw=purple!50!black]
  (-3.85,-0.25)
  .. controls (-2.25,-0.35) and (-0.65,-1.05) .. (0,-2.8)
  .. controls (0.65,-1.05) and (2.25,-0.35) .. (3.85,-0.25);

\draw[thick,samples=700,domain=-4.15:4.15,smooth,draw=blue!50!black,line width=0.9pt]
  plot (\x,{0.155*sin(1450*\x)});

\filldraw (0,2.8) circle (2.2pt) node[right] {\hspace{1.5mm} Equilibrium in the future};
\filldraw (0,-2.8) circle (2.2pt) node[right] {\hspace{1.5mm} Equilibrium in the past};

\node at (-2.5,2.05) {Retarded sector};
\node at (-2.5,1.5) {$\theta(t)$};
\node at (-2.5,-2.05) {Advanced sector};
\node at (-2.5,-1.5) {$\theta(-t)$};

\node at (+3.0,1.5) {Dissipation};
\node at (+3.0,1.0) {$\frac{dS}{dt}>0$};
\node at (+3.0,-1.5) {Dissipation};
\node at (+3.0,-2.0) {$\frac{dS}{dt}>0$};

\draw[->] (-4.2,0.65) -- (-4.2,2.65) node[above] {Asymptotic limit: $t\to\infty$};
\draw[->] (-4.2,-0.65) -- (-4.2,-2.65) node[below] {Asymptotic limit: $t\to-\infty$};

\end{tikzpicture}
}
\vspace{-2mm}

{(a)}
\end{minipage}%
\hfill%
\begin{minipage}{0.5\textwidth}
\centering
\resizebox{\linewidth}{!}{%
\begin{tikzpicture}[>=Latex, line cap=round, line join=round, font=\small]

\tikzset{
  axis/.style={->, line width=0.9pt},
  cont/.style={<->, line width=0.9pt},
  pole/.style={circle, line width=0.9pt, minimum size=4pt,
               fill=blue!50!black, draw=none, inner sep=1.6pt},
  cut/.style={line width=2.4pt, draw=blue!50!black, line cap=butt},
  poles/.style={circle, line width=0.9pt, minimum size=6pt,
               fill=purple!50!black, draw=none, inner sep=1.6pt},
  cutI/.style={line width=2.4pt, draw=purple!50!black, line cap=butt}
}

\def\L{0.5}
\def\R{5}
\def\xmin{-2.1}
\def\xmax{2.1}
\def\ymin{-2.0}
\def\ymax{2.0}

\begin{scope}[shift={(\L,0)}]
  \draw[axis] (\xmin,0) -- (\xmax,0) node[below] {$\mathrm{Re}(z)$};
  \draw[axis] (0,\ymin) -- (0,\ymax) node[above] {$\mathrm{Im}(z)$};

  \node[above left] at (1.5,1.2) {Sheet I};
  \node[above left] at (1.5,0.6) {$G_{A}(z)$};

  \draw[cut] (\xmin,0) -- (\xmax-0.3,0);

  \draw[cut] (-2.05,0.75) -- (-1.7,0.75);
  \node[anchor=west] at (-1.72,0.75) {\scriptsize Branch cut};
\end{scope}

\begin{scope}[shift={(\R+1,0)}]
  \draw[axis] (\xmin,0) -- (\xmax,0) node[below] {$\mathrm{Re}(z)$};
  \draw[axis] (0,\ymin) -- (0,\ymax) node[above] {$\mathrm{Im}(z)$};

  \node[above left] at (1.5,1.2) {Sheet II};
  \node[above left] at (1.5,0.6) {$G^{\mathrm{II}}_{A}(z)$};

  \draw[cutI] (\xmin,0) -- (\xmax-0.3,0);

  \node[poles] at (0.75,-0.75) {};
  \node[poles] at (1.45,-1.05) {};
  \node[poles] at (2.00,-0.65) {};
  \node[poles] at (2.6,-0.9) {};
  \node[poles] at (0.0,0.0) {};
  \node[poles] at (-0.75,-0.75) {};
  \node[poles] at (-1.45,-1.05) {};
  \node[poles] at (-2.00,-0.65) {};
  \node[poles] at (-2.6,-0.9) {};

  \coordinate (EthTwo) at (0.95,-1.0);

  \def\cutlen{1.5}
  \def\ang{-90}
  \draw[cutI] (EthTwo) -- ++(\ang:\cutlen);

    \coordinate (EthTwo) at (-0.95,-1.0);

  \def\cutlen{1.5}
  \def\ang{-90}
  \draw[cutI] (EthTwo) -- ++(\ang:\cutlen);

  \node[poles] at (-1.6,1.15) {};
  \node[anchor=west] at (-1.5,1.15) {\scriptsize Pole};

  \draw[cutI] (-1.80,0.75) -- (-1.5,0.75);
  \node[anchor=west] at (-1.5,0.75) {\scriptsize Branch cut};
\end{scope}


\end{tikzpicture}
}
\vspace{-2mm}

{(b)}
\end{minipage}

\caption{
(a) Asymptotic approach to equilibrium due to the dissipative dynamics contained in mixing systems that possess a branch cut structure in the resolvent of the Liouvillian. The second law of thermodynamics breaks the time symmetry of the solution, requiring entropic increase such that equilibrium is obtained in the future. Therefore, the retarded sector corresponds to the physical solution and the advanced sector can be safely discarded. 
(b) Analytic continuation of $G_{A}(z)$ from the physical sheet (Sheet I) to the second Riemann sheet (Sheet II) $G^{\text{II}}_{A}(z)$ by traveling through the branch cut from above. The invariant measure appears on the second Riemann sheet along with the Ruelle-Pollicott resonances and complex branch cuts. The spectra on both sheets is symmetric about the imaginary axis. 
}
\label{fig:causal_sheets}

\end{figure*}

Combining both these results together, it is straightforward to see that the unitary time evolution group splits into the two semi-groups corresponding to the retarded and advanced sectors: 
\begin{gather}
    \begin{split}
        \hat{U}(t) = \hat{U}_{+}(t) +  \hat{U}_{-}(t) \ .
    \end{split}
\end{gather}
It is important to note that the emergence of semi-group time evolution is also exhibited in interacting quantum systems that have continuous spectra~\cite{antoniou1993intrinsic,antoniou2003irreversibility}. The semi-group structure is clear as $\hat{U}_{\pm}(t)\hat{U}_{\pm}(s)=\hat{U}_{\pm}(t+s) \hspace{1.5mm} \forall\hspace{1mm} t,s \gtrless 0$. Importantly, we have that $\hat{U}_{-}(t)=[\hat{U}_{+}(-t)]^\dag$. As this structure explicitly arises due to the continuous spectrum of $\mathcal{L}$, this necessitates a rigged Hilbert space formulation in order to rigorously construct the eigenstates of $\mathcal{L}$~\cite{bohm2013quantum}. The rigged Hilbert space formulation is also necessary in order to construct the distributional eigenstates of the continuous position and momentum operators in quantum mechanics~\cite{bohm2013quantum,stefanucci2013nonequilibrium}. It is on the larger rigged Hilbert space where the semi-group dissipative nature of the Frobenius-Perron operator is revealed~\cite{gaspard2005chaos,antoniou1993intrinsic,prigogine1987intrinsic,petrosky1991alternative,petrosky1996poincare,petrosky1997liouville,antoniou1993generalized}. This semi-group structure supports forward- and backward-time dissipative evolution involving entropic increase and the approach to equilibrium states. We provide a detailed modern analysis of the extended Banach space construction in Appendices~\ref{app:spec_cont} and~\ref{app:continuation}. In fact, the continuous spectrum of the generator in the thermodynamic limit is deeply connected to the fact that the algebraic state must be associated with a type III von Neumann algebra~\cite{witten2018aps,witten2022does,furuya2023information}. It is clear that the asymptotic limit, $\lim_{t\to\infty} \omega_{t}(A)$, involves only the retarded evolution sector $U_{+}(t)$, giving rise to a stationary equilibrium state in the future, thereby breaking the time-reversible symmetry of the theory (see Fig.~\ref{fig:causal_sheets}(a)). In the context of scattering theory, the dissipative evolution of the eigenstates of the absolutely continuous part of the spectrum is clear from the RAGE theorem: $\lim_{t\to\infty}\braket{\phi|e^{-iHt}|\psi}=0, \hspace{1.5mm}\forall\hspace{0.5mm} \phi,\psi \in \mathcal{H}_{\text{ac}}$~\cite{reed1979iii}.  Importantly, for classical mixing systems, Misra, Prigogine and Courbage showed that it is in fact possible to introduce a time operator~\cite{misra1979deterministic,prigogine1982being} (see Appendix~\ref{app:prigogine}). This means that the internal `age' of the system can be tracked. In the next section, we shall show that for large quantum systems, the time evolution tends to a time-independent equilibrium state (the invariant measure) associated with non-unitary dissipative evolution. Importantly, the semi-group structure emerges naturally from the theory without requiring us to make any modifications.

\section{Emergence of the arrow of time and equilibrium states from the intrinsic dynamics of quantum systems}~\label{sec:mixing_equilibrium}

For mixing systems, where the spectrum of the Liouvillian is absolutely continuous, the invariant measure of the dynamics (stationary equilibrium state) can be obtained from the asymptotic limit of the dynamics:
\begin{gather}
    \begin{split}
        \omega_{t}(A) &=i\int^{\infty}_{-\infty} \frac{d\omega}{2\pi}\  e^{-i\omega t} G_{A}(\omega+i0^{+}) \ ,
    \end{split}
\end{gather}
for $t>0$. 
We can invert the transformation by analytic continuation of the Green's function $G_{A}(z)$ through the branch cut and onto the second Riemann sheet. This allows us to obtain the contour integral for $t>0$ as 
\begin{gather}
    \begin{split}~\label{eq:inv_laplace}
        \omega_{t}(A) = i\int^{\infty}_{-\infty} \frac{d\omega}{2\pi}\ e^{-i\omega t}G^{\text{II}}_{A}(\omega-i0^+) \ ,
    \end{split}
\end{gather}
where we have the analytic continuation identity: $G_{A}(\omega+i0^{+})=G^{\text{II}}_{A}(\omega-i0^+)$. 
This integral now lies on the second Riemann sheet in the lower half of the complex plane. The analytic continuation gives rise to~\cite{gaspard2005chaos,bach2000return} (see Appendix~\ref{app:continuation})
\begin{gather}
    \begin{split}~\label{eq:analy_cont}
        G^{\text{II}}_{A}(z) = \frac{\tilde{\omega}_0(A)}{z}{}+\sum_{J\neq 0}\frac{\tilde{\omega}_{J}(A)}{z-z_{J}} + G^{\text{II,c}}_{A}(z) \ .
    \end{split}
\end{gather}
This structure is depicted in Fig.~\ref{fig:causal_sheets}(b). Due to the fact that the background term and the poles all lie beneath the real axis ($\text{Im}z_{J}<0$), they generate irreversible decay. In Eq.~\ref{eq:analy_cont}, we have assumed that the poles of the analytically continued function are simple, without multiplicity. Here, $G^{\text{II,c}}_{A}(z)$ is the continuum background term of the second Riemann sheet, resulting from the branch cut structure of the analytically continued spectral function. In Appendix~\ref{app:continuation}, we provide a more general analysis of the pole and background branch cut contributions. The analytic continuation is analogous to that employed in $\mathcal{S}$-matrix theory to calculate scattering rates~\cite{reed1979iii,bohm2013quantum,coveney2026thesis}. As the spectrum of $\mathcal{L}$ is absolutely continuous for mixing systems, the analytic continuation of the Green's function reveals a unique simple pole at $z_{0}=0$, as the system is assumed to be mixing with only the energy being conserved by the dynamics. The correspondence with classical systems is striking and the set of poles revealed by the analytic continuation was first realised by Pollicott and Ruelle in the context of classical dynamical systems~\cite{pollicott1985rate,pollicott1986meromorphic,ruelle1986,ruelle1986locating,ruelle1987one,ruelle1989thermodynamic,ruelle1989chaotic}. The existence of quantum Ruelle-Pollicott resonances was also conjectured in Ref.~\cite{prosen2002ruelle} in the context of `chaotic' quantum many-body dynamics. Because the system conserves probability and the total energy, the unique pole at $z_0=0$ must exist. This pole can only be revealed by the analytic continuation as the spectrum of the Liouvillian is absolutely continuous and the stationary equilibrium state must correspond to a generalized distributional eigenstate~\cite{gaspard2005chaos}. In classical dynamical systems theory the corresponding eigenfunction is referred to as the Sinai-Ruelle-Bowen (SRB) measure~\cite{sinai1972gibbs,bowen1975ergodic,ruelle1976measure}. Physically, this measure represents the equilibrium state to which all smooth initial states converge, thereby acting as the attractor for the dynamics.

Using Eq.~\ref{eq:analy_cont}, we can analytically evaluate the integral in Eq.~\ref{eq:inv_laplace} to give 
\begin{gather}
    \begin{split}
         \omega_{t}(A) =  \tilde{\omega}_{0}(A)+\sum_{J\neq 0}e^{-iz_{J}t}\tilde{\omega}_{J}(A) + G^{\text{II,c}}_{A}(t) \ ,
    \end{split}
\end{gather}
with $G^{\text{II,c}}_{A}(t)=i\int^{\infty}_{-\infty} \frac{d\omega}{2\pi}\ e^{-i\omega t}G^{\text{II,c}}_{A}(\omega-i0^+)$.  
Taking the asymptotic limit of this equation, we find 
\begin{gather}
    \begin{split}
        \lim_{t\to\infty}\omega_{t}(A) = \tilde{\omega}_{0}(A) \ .
    \end{split}
\end{gather}
This is because the resonant and background continuum contributions decay to zero, with the resonant states displaying exponential decay while the background continuum term gives rise to longer time tail behaviour such as polynomial decay $\mathcal{O}(t^{-n})$ or more complicated branch cut dependent non-exponential tails $\mathcal{O}(t^{-(\alpha_c+1)}e^{-t})$, where $\alpha_c$ is the complex exponent. The background continuum term vanishes in the asymptotic limit due to the Riemann-Lebesgue theorem. 
From the conservation of probability, we have 
$\tilde{\omega}_{0}(\mathbbm{1})=1$ as well as the positivity condition: $\tilde{\omega}_{0}(A^*A)\geq 0$. If the energy is the only quantity conserved by the dynamics, we also have that the energy is preserved by the state during time evolution: $\omega_{t}(H)=\tilde{\omega}_{0}(H)$. These properties mean that $\tilde{\omega}_{0}$ satisfies the conditions of a state in quantum theory (see Appendix~\ref{app:aqm}) and therefore must correspond to the unique invariant equilibrium state. 
The establishment of a unique equilibrium state is a result of the non-integrability of the mixing dynamics where only the energy is conserved by the motion. The mixing nature of the dynamics means the initial state is always attracted to this invariant measure. For an isolated system, this state corresponds exactly to the microcanonical ensemble~\cite{gaspard2005chaos,petrosky1991alternative}. The state is revealed on the second Riemann sheet as it is a distribution eigenstate and therefore is not an element of the Liouville-Hilbert space. Therefore, we can write
\begin{gather}
    \begin{split}
        \tilde{\omega}_{0} \equiv \omega^{\text{mce}}_{\text{eq}} \ ,  
    \end{split}
\end{gather}
where the state $\omega^{\text{mce}}_{\text{eq}}$ is time-independent and corresponds to the microcanonical ensemble. As a result, the intrinsic dynamics of the system generates the mixed state corresponding to the invariant measure ($\omega^{\text{mce}}_{\text{eq}}$) from an initial pure state: $\omega_{0}$. This involves an increase in entropy of the isolated system in accordance with the second law of thermodynamics (see Appendix~\ref{app:lyapunov} for the construction of the Lyapunov functional and the thermodynamic entropy). It is interesting to note that our thermodynamic entropy construction is similar to that of the Uhlmann-Araki measure~\cite{araki1976relative,uhlmann1977relative,furuya2023information}. In Ref.~\cite{furuya2023information}, it was identified that the existence of the Uhlmann-Araki relative entropy immediately implies the second law. However, the specific dynamical mechanism for its generation was not discussed. Our analysis demonstrates that, for mixing systems, we have the irreversible generation of an equilibrium state from the intrinsic dynamics of the system:
\begin{gather}
\begin{split}
   \lim_{t\to\infty}\omega_{t}(A) = \omega^{\text{mce}}_{\text{eq}}(A) \ .
\end{split}
\end{gather}
It is important to note that this asymptotic state has emerged from the underlying microscopically reversible dynamics of the Liouville equation. However, the arrow of time \emph{emerges} as a result of a macroscopically large number of particles interacting with each other which gives rise to the broken time-symmetry of the solutions~\cite{prigogine2018order,coveney1991arrow}. In this way, the intrinsic dynamics of an interacting many-particle system acquires a direction of time. It is important to note that if there are other conserved charges $Q_i$, such that $\mathcal{L}Q_{i}=0$, the invariant equilibrium state will reflect the conservation of these quantities and will not correspond to a unique pole representing the microcanonical ensemble.

It is important to note that the relation $\hat{U}_{-}(t) = [\hat{U}_{+}(-t)]^\dag$, means that the backward-time advanced asymptotic limit also exists:
\begin{gather}
    \begin{split}
        \lim_{t\to-\infty}\omega_{t}(A) = \omega^{\text{mce}}_{\text{eq}}(A) \ ,
    \end{split}
\end{gather}
with entropic increase and the equilibrium state being obtained in the past. The invariant measure reached by backward-time evolution is the same as that reached by the forward-time evolution due to the fact that the equations of the system are time-reversal symmetric. The general property of symmetry-broken solutions is that they appear in pairs. Therefore, the time-inversion symmetric equations yield solutions that come in pairs, one corresponding to the retarded-time direction and the other to the advanced-time direction. However, the advanced-time direction is in direct contradiction with the second law of thermodynamics which requires the entropy of the universe to increase with time. This is ultimately the consequence of the fact that the universe began in a low entropy state. It means that the advanced solution is in direct contradiction with physical reality and must be discarded. This is analogous to the boundary conditions necessary to obtain the correct physical solution to a given differential equation. As a consequence the time-reversal symmetry is broken, leading to the physical forward-time semi-group evolution corresponding to physical reality in agreement with the second law. Hence, quantum mechanical systems that are mixing intrinsically approach equilibrium in the future, breaking the time-reversal symmetry of the microscopic dynamics leading to the irreversible increase in entropy. We emphasise here that we have demonstrated this for an isolated quantum system. The generation of broken symmetry phases is widespread in condensed matter physics, for example macroscopic magnetic phases acquire a non-zero macroscopic magnetization as a result of the thermodynamic limit, with the solutions appearing in pairs $(\pm M)$~\cite{anderson1958absence,anderson1961localized,anderson1963plasmons,anderson1972more,altland2010condensed}. Indeed, spontaneous symmetry-breaking is also commonly seen in high energy physics. For example, the Higgs mechanism leads to the vacuum itself not being perfectly symmetrical~\cite{higgs1964broken,englert1964broken}; other examples include CP violation which is the cause of matter/anti-matter imbalance. In addition, reaction-diffusion systems can also lead to physical behaviour that breaks the spatial inversion symmetry of the equations~\cite{prigogine2018order}.  

The complex eigenvalues appearing in Eq.~\ref{eq:analy_cont} can be related to the eigenvalues of the analytically continued Liouvillian discussed by Petrosky and Prigogine~\cite{petrosky1991alternative,petrosky1996poincare,petrosky1997liouville}. However, their approach was based around the Hilbert-Schmidt formulation of Liouville space which does not allow for a correct definition of the microcanonical distribution as the identity operator does not exist in the Hilbert space in the thermodynamic limit (because then $\text{Tr}\{\mathbbm{1}\}$ is unbounded). Thus their formalism does not rigorously account for the approach to equilibrium or the thermodynamic limit of states and observables. In order to overcome these limitations, our approach is formulated in terms of algebraic quantum states, which allows for the description of infinite systems such as those defined in the thermodynamic limit or quantum fields~\cite{witten2022does} (see Appendices~\ref{app:aqm},~\ref{app:spec_cont} and~\ref{app:continuation}). In addition, as we have demonstrated, the mixing dynamics of the quantum system is intrinsic to the theory itself and there is no need to introduce any `star-unitary' transformations as a modification of the fundamentally reversible LvN equation~\cite{prigogine1982being}. 

The relationship between the complex representation of Eq.~\ref{eq:analy_cont} and the necessity of an absolutely continuous spectrum of $\mathcal{L}$ was also rigorously demonstrated by Bach, Fröhlich and Sigal~\cite{bach2000return}. However, their work was based on the study of an open quantum subsystem coupled to an infinite thermodynamic heat bath which is already initially in equilibrium.
But their work fails to explain how a quantum system reaches a state of equilibrium in the first place. In addition, their analysis also shows that `return to equilibrium' for the open quantum system is obtained both in the future (retarded direction) as well as in the past (advanced direction). As we have stated, this clearly indicates the presence of  time symmetry broken solutions, with the second law providing the selection principle for the physical forward-time evolution. However, this explicit symmetry breaking of the solutions was not identified in their work~\cite{bach2000return}.

It is interesting to note that quantum Ruelle-Policott resonances have also been uncovered in the context of black hole studies involving quantum gravity with connections to holography~\cite{polchinski2015chaos,kapec2023photon,dodelson2025black}. In particular, they have been identified in the study of `quasinormal' modes of a perturbed black hole as it approaches an equilibrium state. This correspondence was identified due to the fact that that black holes are thought to behave much like isolated quantum systems that exhibit extremely chaotic dynamics.

Moreover, the approach to equilibrium is not possible for finite and small quantum systems as the asymptotic limit does not exist: $\lim_{t\to\infty}\omega_{t}(A)\neq\omega^{\text{mce}}_{\text{eq}}(A)$. This is because there is no dissipation in the dynamics as the resolvent of the Liouvillian is simply a meromorphic function. This means that the system evolves unitarily forever. However, this is not observed for many macroscopically large quantum systems which are described by the thermodynamic limit. Such oscillatory behaviour is effectively of zero measure for any realistic “interacting” quantum system. This is analogous to the fact that finite systems do not exhibit phase transitions, demonstrating the importance of the thermodynamic limit in the treatment of quantum mechanical systems involving large numbers of interacting particles.

In addition, our work provides a dynamical explanation for the Eigenstate Thermalization Hypothesis (ETH)~\cite{deutsch1991quantum,srednicki1994chaos,d2016quantum,hahn2024eigenstate,capizzi2025hydrodynamics,ippoliti2026infinite,serbyn2026eigenstate}. The ETH states that every energy eigenstate of a `chaotic' quantum system is already in a thermal state such that the expectation value of an observable is exactly given by that of the microcanonical ensemble average. However, this proposition is not derived from the intrinsic dynamics corresponding to a macroscopically large non-integrable quantum system. Our work demonstrates the validity of this hypothesis as we show that the intrinsic dynamics of a mixing system, where only energy is conserved, possesses an  attractor for the dynamics that is exactly the microcanonical invariant measure.

\section{Application to quantum measurement and irreversible wave function collapse}~\label{sec:measurement}

The approach we take here is philosophically similar to that which motivated the many-worlds interpretation of quantum mechanics~\cite{everett1957hugh}: quantum measurement is a dynamical process that must also be described by quantum theory. However, we shall demonstrate that a more sophisticated analysis of the mixing dynamics of quantum measurement gives rise to objective probabilistic wave function collapse displayed for the entire isolated quantum system. 

To apply our formalism to the quantum mechanical measurement process, we employ a two-state spin-boson model. The measurement apparatus is represented by a macroscopic quantum system and for simplicity we take a two state model representing the two different projections of a spin. The Hamiltonian is written as 
\begin{gather}
    \begin{split}~\label{eq:measure_ham}
        H &= \sum_{i=\{\uparrow,\downarrow\}}\epsilon_{i} a^\dag_{i}a_{i} + \int_{\mathbbm{R}^3} d\mathbf{k}\ \omega(\mathbf{k}) b^\dag_\mathbf{k}b_\mathbf{k}\\
        &+ \sum_{i=\{\uparrow,\downarrow\}}\int_{\mathbbm{R}^3} d\mathbf{k}\ g_{i}(\mathbf{k}) A_{\mathbf{k}} a^\dag_{i}a_{i} \\
        &+\frac{1}{2}\int_{\mathbbm{R}^3} d\mathbf{k}d\mathbf{q} d\mathbf{p}\ v(\mathbf{p})b^\dag_{\mathbf{k-p}}b^\dag_{\mathbf{q+}\mathbf{p}}b_{\mathbf{q}}b_{\mathbf{k}}\ ,
    \end{split}
\end{gather}
where $A_\mathbf{k}=b_\mathbf{k}+b^\dag_{-\mathbf{k}}$ represents the macroscopic displacement of the measuring apparatus that is coupled to the subsystem (spin projections) caused by the measurement process. This Hamiltonian represents an ideal measurement process outlined initially by von Neumann~\cite{von2018mathematical}. Crucially, this means that $[H,a^\dag_{i}a_{i}] = 0 $ and the projection onto the spin manifold is conserved by the dynamics.  As first formalized by Bohm, this type of interaction is what allows the measuring apparatus to pick out one of the basis states with a definite probability~\cite{bohm2012quantum}. Importantly, for the macroscopic measuring device to register or distinguish between the two states requires $g_{\uparrow}(\mathbf{k})\neq g_{\downarrow}(\mathbf{k})$. The final term of Eq.~\ref{eq:measure_ham} represents the generic momentum transferring interaction between the different measurement modes. We could also include other higher-order anharmonic interactions between the measurement modes but we choose to write only the two-body interaction for simplicity. The spectrum of the non-integrable Hamiltonian of Eq.~\ref{eq:measure_ham} is absolutely continuous due to the coupling to the thermodynamic measurement apparatus. This is similar to the Friedrichs model, which also contains an absolutely continuous spectrum, providing a useful model of quantum mechanical decay and the emergence of irreversible time-evolution~\cite{friedrichs1948perturbation,prigogine1987intrinsic,antoniou1993intrinsic}. The second quantized field operators for the basis states are defined as $a^\dag_{\uparrow}\ket{\text{vac}}=\ket{\uparrow}$ and $a^\dag_{\downarrow}\ket{\text{vac}}=\ket{\downarrow}$, respectfully. The initial subsystem wave function is given by $\ket{\psi^{s}_0}=\alpha\ket{\uparrow}+\beta\ket{\downarrow}$, with $|\alpha|^2+|\beta|^2=1$.

First, we demonstrate that the reduced density matrix of the 
spin subsystem evolves to a mixed state. The central quantity we are concerned with is the reduced one-particle density matrix of the spin subsystem given by 
\begin{gather}
    \begin{split}
        \rho^{s}_{ij}(t) = \braket{P_{ij}(t)} = \braket{a^\dag_{i}(t)a_{j}(t)} \ ,
    \end{split}
\end{gather}
which is the time-dependent projector onto the basis states being measured. The expectation value is taken over the initial state of the total isolated system.  
The equation-of-motion of the reduced density matrix of the subsystem is given by 
\begin{gather}
    \begin{split}
        \left(i\frac{\partial}{\partial t}-\Delta_{ij}\right)\rho^{s}_{ij}(t) = \int_{\mathbbm{R}^3} d\mathbf{k} \Big(g_{i}(\mathbf{k})-g_{j}(\mathbf{k})\Big)\braket{A_\mathbf{k}(t)P_{ij}(t)} ,
    \end{split}
\end{gather}
with $\Delta_{ij}=\epsilon_{i}-\epsilon_{j}$. As the equation-of-motion does not mix the different basis components of the density matrix, we can immediately close this expression by introduction of the multiplicative self-energy which gives rise to the Volterra equation (see Appendix~\ref{app:gen_dm})
\begin{gather}
    \begin{split}
       &\int_{\mathbbm{R}^3} d\mathbf{k} \Big(g_{i}(\mathbf{k})-g_{j}(\mathbf{k})\Big)\braket{A_\mathbf{k}(t)P_{ij}(t)} \\
       &=  \int^{t}_{0} d\Bar{t}\  \Pi_{ij}(t-\Bar{t})\rho^{s}_{ij}(\Bar{t})  .
    \end{split}
\end{gather}
This allows us to formally close the equation-of-motion as 
\begin{gather}
    \begin{split}
        \left(i\frac{\partial}{\partial t}-\Delta_{ij}\right)\rho^{s}_{ij}(t) =  \int^{t}_{0} d\Bar{t}\  \Pi_{ij}(t-\Bar{t})\rho^{s}_{ij}(\Bar{t}) \ .
    \end{split}
\end{gather}
Taking the Laplace transformation of this expression, we have 
\begin{gather}
    \begin{split}
        \rho^s_{ij}(z) = G_{ij}(z)\rho^{s,0}_{ij} \ , 
    \end{split}
\end{gather}
where $\rho^{s,0}_{ij}=\rho^s_{ij}(t=0)$ is the initial reduced density matrix and the Green's function is given by 
\begin{gather}
    \begin{split}
        G_{ij}(z) = \frac{1}{z-\Delta_{ij}-\Pi_{ij}(z)} \ .
    \end{split}
\end{gather}
Inverting the transformation, we have 
\begin{gather}
    \begin{split}
        \rho^s_{ij}(t) = G_{ij}(t) \rho^{s,0}_{ij}\ .
    \end{split}
\end{gather}
The exact properties of the self-energy of this model can be immediately inferred: $\Pi_{ii}(z) = 0$, $\Pi_{ij}(z) \neq 0$ and $\text{Im}\Pi_{ij}(z) \neq 0$ for $i\neq j$. The non-zero imaginary part of the off diagonal self-energy entirely results from the coupling to the macroscopic measurement apparatus. In fact, the Laplace transform of the reduced density matrix is closely related to the projection of the full resolvent: $\braket{P^\dag_{ij}(z\mathbbm{1}-\mathcal{L})^{-1}P_{ij}}$. Therefore, the branch cut structure of the self-energy is directly inherited from the branch cut structure of the full resolvent. This means that the irreversibility of the dynamics in the projected sector of the reduced density matrix represents the shadow of the overall irreversibility of the dynamics of the total isolated system. This is only possible in the thermodynamic limit where the spectrum of the non-integrable Liouvillian is absolutely continuous. From the vanishing diagonal components of the self-energy, we immediately have 
\begin{gather}
    \begin{split}~\label{eq:eqm_cond}
        \rho^s_{ii}(z) =  \frac{\rho^{s,0}_{ii}}{z} \ ,
    \end{split}
\end{gather}
using $\Delta_{ii} = 0$. Inverting this relationship through the Bromwich integral, we find the correct conservations of the spin projections:
\begin{gather}
    \begin{split}
        \rho^s_{ii}(t) = \rho^{s,0}_{ii} \ .
    \end{split}
\end{gather}
Using analytic continuation, the propagator can be expressed on the second Riemann sheet as the sum over complex poles and background branch contributions. This is reflective of the fact that the first sheet Green's function contains a branch cut due to the continuous spectrum of the Liouvillian. Therefore, the exact Laplace solution is 
\begin{gather}
    \begin{split}~\label{eq:cont_rep}
        \rho^s_{ij}(z) = \left(\frac{R^{\text{eq}}_{ij}}{z}+\sum_{J\neq 0}\frac{\tilde{R}^{J}_{ij}}{z-z_{J}}+ G^{\text{II,c}}_{ij}(z)\right)\rho^{s,0}_{ij} \ , 
    \end{split}
\end{gather}
with $\text{Im}z_{J}\leq 0$ from the causality conditions and where $\tilde{R}^{J}_{ij}=\left(1-\frac{\partial\Pi_{ij}}{\partial z}\Big|_{z=z_{J}}\right)^{-1}$ are the complex residues. From Eq.~\ref{eq:eqm_cond} we know that $R^{\text{eq}}_{ii}=1$, with $\tilde{R}^{J}_{ii}=G^{\text{II,c}}_{ii}(z)=0$. Eq.~\ref{eq:cont_rep} immediately gives the real-time reduced density matrix as 
\begin{gather}
    \begin{split}~\label{eq:exact_solution}
        \rho^s_{ij}(t) = F_{ij}(t)\rho^{s,0}_{ij} \ , 
    \end{split}
\end{gather}
where $F_{ij}(t) = R^{\text{eq}}_{ij}+\sum_{J\neq 0}\tilde{R}^{J}_{ij}e^{-iz_{J}t}+ G^{\text{II,c}}_{ij}(t)$. Importantly, Eq.~\ref{eq:exact_solution} preserves the hermiticity of the density matrix, its trace and the positivity of its eigenvalues. 
As the individual spin projections are conserved,  we can immediately identify $R^{\text{eq}}_{ij}=0$ when $i\neq j$ which follows from the pure dephasing interaction. Therefore, we can write 
\begin{gather}
    \begin{split}
        \rho^s_{ij}(t) = \left(\sum_{J\neq 0}\tilde{R}^{J}_{ij}e^{-iz_{J}t}+ G^{\text{II,c}}_{ij}(t)\right)\rho^{s,0}_{ij} \ ; \hspace{2mm}i\neq j \ .
    \end{split}
\end{gather}
In matrix form, using the expression for the initial reduced density matrix $\rho^{s,0}_{ij}$, we have 
\begin{gather}
    \begin{split}
        \mathbf{\rho}_s(t) = \left(\begin{array}{cc}
        |\alpha|^2 & \Bar{\alpha}\beta F_{\uparrow\downarrow}(t) \\
        \Bar{\beta}\alpha \overline{F_{\uparrow\downarrow}(t)} & |\beta|^2
    \end{array}\right) \ .
    \end{split}
\end{gather}
We know from the analytic structure and continuation of the resolvent that the time-dependent dissipation will contain long-time tails but ultimately tend to zero ($\lim_{t\to\infty}\rho^s_{\uparrow\downarrow}(t)=0$) such that the asymptotic solution is given by 
\begin{gather}
    \begin{split}
        \lim_{t\to\infty}  \mathbf{\rho}_s(t) = \left(\begin{array}{cc}
        |\alpha|^2 & 0 \\
        0 & |\beta|^2
    \end{array}\right) = \rho^{\text{eq}}_s \ .
    \end{split}
\end{gather}
Therefore, we have obtained an equilibrium solution for the subsystem which corresponds to a mixed state from an initial state that was pure. This process is fundamentally irreversible and directly the result of coupling to the macroscopic measurement apparatus. As shall be explained, the exact approach presented here for the evolution of the reduced density matrix is fundamentally different to that taken in quantum decoherence theory. To summarize in terms of the basis states of the spin subsystem, we find that the initial density operator has gone from the pure state 
\begin{gather}
    \begin{split}
        \rho^{0}_{s} = |\alpha|^2\ket{\uparrow}\bra{\uparrow} + \alpha\Bar{\beta} \ket{\uparrow}\bra{\downarrow} + \bar{\alpha}\beta \ket{\downarrow}\bra{\uparrow} + |\beta|^2\ket{\downarrow}\bra{\downarrow}
    \end{split}
\end{gather}
to the mixed state 
\begin{gather}
    \begin{split}~\label{eq:sys_rho_eq}
        \rho^{\text{eq}}_{s} = |\alpha|^2\ket{\uparrow}\bra{\uparrow} + |\beta|^2\ket{\downarrow}\bra{\downarrow} \ .
    \end{split}
\end{gather}
As a result, we find that the von Neumann entropy of the subsystem increases: $S^{\text{eq}}_{s}> S^{0}_{s}$, where 
\begin{subequations}
    \begin{align}
        \begin{split}
        S^{\text{eq}}_s &= -k_B\text{tr}\{\rho^{\text{eq}}_{s}\text{log}\rho^{\text{eq}}_s\}
        \end{split}\\
        \begin{split}
         S^{0}_s &= -k_B\text{tr}\{\rho^{0}_s\text{log}\rho^{0}_s\} \ .
    \end{split}
    \end{align}
\end{subequations}
This is in direct agreement with the second law of thermodynamics which determines the time direction of the asymptotic time evolution as $\Delta S > 0$. In Appendices~\ref{app:lyapunov} and~\ref{app:gen_dm}, we discuss in further detail the relationship between the evolution of the entropy of the isolated system and the subsystem. Importantly, this is the result of irreversible time evolution of the non-integrable isolated system which in turn is the consequence of the continuous spectrum due to the thermodynamic limit. 
The result is that the Born rule is directly obtained from the underlying dynamics. For example, the expectation value corresponding to an observable of the subsystem $A_s$, can be written as 
\begin{gather}
    \begin{split}
       \omega_{\text{eq}}(A_{s}) &= \sum_{ij}a_{ij}\rho^{s,\text{eq}}_{ji} \\
        &= |\alpha|^2 a_{\uparrow\uparrow} + |\beta|^2 a_{\downarrow\downarrow} \ ,
    \end{split}
\end{gather}
which is a classical probability distribution giving the result of the measurement of $a_{\uparrow\uparrow}$ with probability $p_{\uparrow}=|\alpha|^2$ and $a_{\downarrow\downarrow}$ with probability $p_{\downarrow}=|\beta|^2$. In reality, a given result occurs that is intrinsically random. For example, after measurement of $z$-axis projection, the probability of measuring a given x-axis projection is 
\begin{gather}
    \begin{split}
        \omega_{\text{eq}}(\sigma_{x}) &= |\alpha|^2 \sigma^{x}_{\uparrow\uparrow} + |\beta|^2 \sigma^{x}_{\downarrow\downarrow} \\
        &= 0 \ ,
    \end{split}
\end{gather}
in exact agreement with standard single-particle quantum mechanics. However, this analysis is not enough to explain why a single irreversible outcome occurs for the full isolated system. We now need to analyse the evolution of the full isolated system.

For mixing non-integrable systems where the spectrum of the Liouvillian is continuous, as demonstrated in the previous section, the approach to equilibrium is irreversible for the entire isolated system such that there is a universal entropic increase resulting from the quantum measurement. This is an objective outcome and is not the result of arbitrary coarse-graining approximations.

To demonstrate that this is the case, we provide an analysis of the evolution of the state of the total isolated system consisting of the spin subsystem and the thermodynamic measuring apparatus. The resolvent of the mixing Liouvillian corresponding to Eq.~\ref{eq:measure_ham} contains a branch cut across the entire real axis. This gives rise to the broken time-symmetry in the dynamics and the emergence of intrinsic retarded and advanced time sectors, with the retarded sector corresponding to physical reality. We also know that the spin projections commute with the total Hamiltonian such that they are conserved by the dynamics. In this case, the equilibrium state must also include the conserved `charges' corresponding to the initial spin projections of the spin subsystem: $[H,P_{ii}]=0$. This means that the $z=0$ pole of the analytically continued resolvent of the Liouvillian must be two-fold degenerate (as there are two spin projections) corresponding to the asymptotic equilibrium state (SRB measure). This arises as the Hamiltonian decomposes into two sectors, corresponding to the conservation of the different spin projections: $H=H_{\uparrow}\oplus H_{\downarrow}$. Using the analysis presented in the previous section, the dynamics of the state is given by solution of the inverse Laplace transform
\begin{gather}
    \begin{split}
        \omega_{t}(A) = i\int^{\infty}_{-\infty} \frac{d\omega}{2\pi}\ e^{-i\omega t}G^{\text{II}}_{A}(\omega-i0^+) \ ,
    \end{split}
\end{gather}
where 
\begin{gather}
    \begin{split}
        G^{\text{II}}_{A}(z) = \frac{1}{z}\left(\sum_{i=\{\uparrow,\downarrow\}}p_{i}\omega^{\text{eq}}_{i}(A)\right)+\sum_{J\neq 0}\frac{\tilde{\omega}_{J}(A)}{z-z_{J}} + G^{\text{II,c}}_{A}(z)
    \end{split}
\end{gather}
is given by the analytic continuation of the resolvent of the Liouvillian through the branch cut and onto the second Riemann sheet. The resulting solution is given by  
\begin{gather}
    \begin{split}
        \omega_{t}(A) = \sum_{i=\{\uparrow,\downarrow\}}p_{i}\omega^{\text{eq}}_{i}(A)+\sum_{J \neq 0}e^{-iz_{J}t}\tilde{\omega}_{J}(A) + G^{\text{II,c}}_{A}(t) \ .
    \end{split}
\end{gather}
The asymptotic time evolution of the state of the full isolated system is therefore obtained as 
\begin{gather}
    \begin{split}
    \lim_{t\to\infty}\omega_{t}(A) = p_{\uparrow}\omega^{\text{mce}}_{\uparrow}(A) + p_{\downarrow}\omega^{\text{mce}}_{\downarrow}(A) \ , 
    \end{split}
\end{gather}
with $p_{\uparrow}=|\alpha|^2$ and $p_{\downarrow}=|\beta|^2$. 
These probabilities appear in the equilibrium state as they are constants of the motion. As the spin projections along with the total energy of the system are conserved by the dynamics, the equilibrium distribution is still of microcanonical form but splits into different sectors involving the conservation of the spin projection where $E_{\text{tot}}=E_{\uparrow}+\epsilon_{\uparrow}=E'_{\downarrow}+\epsilon_{\downarrow}$, corresponding to the different energies the spin projection can take in the subsystem being measured. Therefore, the total equilibrium state of the isolated system is the convex combination of states
\begin{gather}
    \begin{split}~\label{eq:total_rho}
        \omega^{\text{tot}}_{\text{eq}} = p_{\uparrow}\omega^{\text{mce}}_{\uparrow} + p_{\downarrow}\omega^{\text{mce}}_{\downarrow} \ .
    \end{split}
\end{gather}
This decomposition ensures that $\omega^{\text{tot}}_{\text{eq}}(P_{ii})=p_{i}$ conserves the probability of the spin projections as well as the total probability: $\omega^{\text{tot}}_{\text{eq}}(\mathbbm{1})=p_{\uparrow}+p_{\downarrow}=1$. To demonstrate that Eq.~\ref{eq:total_rho} recovers the same reduced density matrix as that given by Eq.~\ref{eq:sys_rho_eq}, we take the observable $A_{s}$ to correspond to that of the subsystem only. In this case, we have 
\begin{gather}
    \begin{split}
        \omega^{\text{tot}}_{\text{eq}}(A_s) &= p_{\uparrow}\omega^{\text{mce}}_{\uparrow}(A_s) + p_{\downarrow}\omega^{\text{mce}}_{\downarrow}(A_s)\\
        &= p_{\uparrow}\omega^{\text{eq}}_{s,\uparrow}(A_s) + p_{\downarrow}\omega^{\text{eq}}_{s,\downarrow}(A_s) \ .
    \end{split}
\end{gather}
This equality holds as $\omega^{\text{mce}}_{\uparrow}(A_s)=\omega^{\text{eq}}_{s,\uparrow}(A_s)$, where the state $\omega^{\text{eq}}_{s,\uparrow}$ depends only on the subsystem degrees of freedom.  
This can always be written as the trace over the reduced density matrix of the subsystem such that 
\begin{gather}
    \begin{split}
        \omega^{\text{tot}}_{\text{eq}}(A_s) = \text{tr}\{\rho^{\text{eq}}_{s}A_s\} \ ,
    \end{split}
\end{gather}
where the trace runs over the subsystem degrees of freedom $\{\uparrow,\downarrow\}$. This gives 
\begin{gather}
    \begin{split}
        \text{tr}\{\rho^{\text{eq}}_{s}A_s\} = p_{\uparrow}\omega^{\text{eq}}_{s,\uparrow}(A_s) + p_{\downarrow}\omega^{\text{eq}}_{s,\downarrow}(A_s)
    \end{split}
\end{gather}
such that 
\begin{gather}
    \begin{split}
        \rho^{\text{eq}}_{s} = p_{\uparrow}\ket{\uparrow}\bra{\uparrow} + p_{\downarrow}\ket{\downarrow}\bra{\downarrow} \ .
    \end{split}
\end{gather}
This expression is exactly identical to that given in Eq.~\ref{eq:sys_rho_eq}. Therefore, we have shown that the total equilibrium state of the full isolated system is given by the two-fold degenerate sectors corresponding to the conserved spin projections. It is crucial to note that the entropy of both the subsystem as well as the entire isolated system has increased as a result of the measurement interaction giving rise to stationary states. This is because both the state of the subsystem and the full isolated system have evolved from initial pure states to mixed states. This is exactly in agreement with the entropic increase first postulated by von Neumann~\cite{von2018mathematical}. 

We have demonstrated that the isolated system of the measurement Hamiltonian (Eq.~\ref{eq:measure_ham}) evolves to a mixed equilibrium state that conserves the projection of the quantity being measured in the subsystem, with the entropy of the total isolated system increasing in accordance with the second law. Eq.~\ref{eq:total_rho} shows that the equilibrium state can be represented as a combination of stationary states which preserve the conserved spin projections. In reality quantum mechanical measurement gives rise to a definite outcome of the corresponding observable. However, the equilibrium states $\omega^{\text{mce}}_{\uparrow}$ and $\omega^{\text{mce}}_{\downarrow}$ are in fact disjoint extremal states corresponding to fundamentally different macroscopic realities that cannot  communicate or exchange energy. This is analogous to the phenomenon of symmetry breaking of thermodynamic phases of magnetic materials~\cite{anderson1958absence,anderson1961localized,anderson1963plasmons}. Therefore, wave function collapse into a definite state is a result of the intrinsic dynamics in the thermodynamic limit that selects an extremal disjoint equilibrium state, with the probability of the system collapsing into the given macroscopic reality given by the weight of the conserved coefficients. Importantly, the coefficients corresponding to the distinct thermodynamic states are not arbitrary but are determined by the initial wave function of the subsystem that is preserved in a pure measurement~\cite{bohm2012quantum}. This is the dynamical origin of the Born rule~\cite{born1926quantenmechanik}. The final states correspond to macroscopically different pointer states (either up or down for the spin subsystem) and therefore must correspond to distinct, mutually exclusive realities. By re-writing the measurement interaction as 
\begin{widetext}
    \begin{gather}
    \begin{split}
         \sum_{i=\{\uparrow,\downarrow\}}\int_{\mathbbm{R}^3} d\mathbf{k}\ g_{i}(\mathbf{k}) A_{\mathbf{k}} a^\dag_{i}a_{i} &=\int_{\mathbbm{R}^3} d\mathbf{k}\ g_{o}(\mathbf{k}) A_{\mathbf{k}}  \sum_{i=\{\uparrow,\downarrow\}}a^\dag_{i}a_{i}+ \int_{\mathbbm{R}^3} d\mathbf{k}\ \Delta g(\mathbf{k}) A_{\mathbf{k}} \left(a^\dag_{\uparrow}a_{\uparrow}-a^\dag_{\downarrow}a_{\downarrow}\right) ,
    \end{split}
\end{gather}
\end{widetext}
where $g_{o}(\mathbf{k})=\frac{1}{2}(g_{\uparrow}(\mathbf{k})+g_{\downarrow}(\mathbf{k}))$ and $\Delta g(\mathbf{k})=\frac{1}{2}(g_{\uparrow}(\mathbf{k})-g_{\downarrow}(\mathbf{k}))$, we can define the thermodynamic pointer observable 
\begin{gather}
    \begin{split}
        M = \int_{\mathbbm{R}^3} d\mathbf{k}\ \Delta g(\mathbf{k}) A_{\mathbf{k}} \ ,
    \end{split}
\end{gather}
which is coupled directly to the measured projection through: $M\otimes(a^\dag_\uparrow a_{\uparrow}-a^\dag_\downarrow a_{\downarrow})$. As the Hamiltonian splits into two sectors that preserve the spin projections, we have 
\begin{subequations}
    \begin{align}
        \begin{split}
            \omega^{\text{mce}}_{\uparrow}(M) &= m_{\uparrow}
        \end{split}\\
        \begin{split}
             \omega^{\text{mce}}_{\downarrow}(M) &= m_{\downarrow} \ ,
        \end{split}
    \end{align}
\end{subequations}
where $m_{\uparrow}\neq m_{\downarrow}$ by construction of the measurement interaction. Therefore, the macroscopic pointer observable $M$ distinguishes the two equilibrium states. By definition, a measurement interaction must ensure that the values obtained from the macroscopic pointer observable are distinct and well defined. Consequently, the two equilibrium states $\omega^{\text{mce}}_{\uparrow}$ and $\omega^{\text{mce}}_{\downarrow}$ correspond to different macroscopic values of the same pointer observable and therefore belong to disjoint thermodynamic states. The fact that initially coherent quantum states can give rise to disjoint thermodynamic states as a result of quantum measurement was first proven by Hepp~\cite{Hepp1972}. However, our work provides the fundamental dynamical explanation for the establishment of these disjoint states which occur as the result of irreversible evolution that leads to entropic increase, both important issues that Hepp acknowledged were missing from his approach~\cite{Hepp1972}. 

Every realized macroscopic state is an extremal phase and the Born rule in fact corresponds to the conserved spectral weights that define the probability measure of the degenerate equilibrium manifold ($z=0$) corresponding to the set of realizable thermodynamic states. The quantum measurement problem has therefore been reduced to the asymptotic equilibrium attractor state of the intrinsic dynamics due to interaction between the macroscopic quantum measuring apparatus and the quantum subsystem. This result is much stronger than decoherence which only applies to the reduced density matrix of the subsystem and does not establish the fact that the dynamics of the full isolated system is fundamentally irreversible. In addition, it is not clear at all from the reduced density matrix how a distinct outcome is realized. This is because the reduced density matrix simply represents the shadow of the global equilibrium state and is not a fundamental quantity. The total global thermodynamic state decomposes into disjoint macroscopic states corresponding to the different outcome of measurement
observables. The Born weights naturally appear as the coefficients conserved by the dynamics in the decomposition of the system into mixed states. The full global thermodynamic states of the isolated system are necessary in order to properly interpret the measurement outcome. The measurement interaction has transformed what would have been a microscopic quantum state into two completely distinct, mutually exclusive macroscopic worlds. This results in the inherently probabilistic objective collapse of the wave function into a distinct macroscopic state given by the conserved Born rule at the time of measurement: $p_{n}=|c_{n}|^2$.

It is important to emphasise the differences between the theory presented in this paper and the notion of quantum decoherence theory~\cite{zurek2003decoherence}. In contrast with the latter, our work establishes irreversible time evolution of a macroscopic isolated quantum system in the thermodynamic limit. This takes place at the level of both the reduced density matrix as well as the quantum state of the isolated system. Our approach is conceptually simple: we treat the dynamics of the full isolated macroscopic quantum state (including system and environment) using the physically correct thermodynamic limit. As we have demonstrated, this allows us to treat a quantum system as isolated, without having to refer to the \emph{ad hoc} definition of an `open system', leading to time symmetry breaking, the emergence of the arrow of time, irreversibility, the reconciliation of quantum dynamics with the second law of thermodynamics and collapse of the wave function thereby providing the dynamical origin of the Born rule. In the process, we also demonstrate that the approach to equilibrium gives the correct form of the microcanonical ensemble. These foundational results are all in agreement with experimental observation. We end up with probabilistic classical outcomes as the intrinsic quantum dynamics of the measurement process establishes disjoint thermodynamic states that can no longer interact with each other~\cite{Hepp1972}. 

Quantum decoherence theory, however, is formulated based on the principle that a quantum system can never be isolated from its environment. This assertion has paved the way for the commonly held assumption that macroscopic quantum systems are fundamentally `open quantum systems'. In this way, a quantum system is modelled as always being coupled to a macroscopic environment, thereby leading to its decoherence by interaction with the environment. This deftly sidesteps the issues of dissipative evolution, the approach to equilibrium and wave function collapse as it merely moves the goalposts by attributing these issues in some unspecified manner to `the environment'. This is because the main purpose of this theory is to preserve the unitary evolution of the overall system and environment, such that time symmetry is maintained and wave function collapse is eliminated from consideration. However, quantum decoherence is unable to explain the establishment of equilibrium states in isolated and closed quantum systems, let alone the open systems it purports to describe. The open systems approach cannot be used to explain the origin of any quantum kinetic equations, such as the Boltzmann
equation, as these are designed to describe the approach to equilibrium states which do not exist according to the model. In Appendices~\ref{app:lyapunov} and~\ref{app:gen_dm}, we provide a detailed analysis demonstrating how the second law of thermodynamics can be recovered from the intrinsic dynamics of macroscopic quantum systems, thereby explaining the monotonic increase in the entropy of an isolated system and the non-monotonic increase in the entropy of an embedded subsystem. In contrast, decoherence theory states that there is no such entropy change as there is no time-symmetry breaking in the dynamics and is constrained to explain the observed increase in the entropy of the subsystem in terms of an equal but opposite change in the entropy of `the environment'. As our analysis shows, the explanations provided by decoherence theory are not sufficient since the intrinsic dynamics of macroscopic quantum systems and the second law of thermodynamics are entirely reconcilable.

Furthermore, in order to extract answers from a quantum computation requires many measurements to be performed. This means that a quantum computer is actually a macroscopic device attached to the quantum hardware. In quantum decoherence theory, the wave function is said to have decohered but no measurement outcome is mentioned. Our work instead paves the way toward a dynamical analysis of the quantum measurement process that can lead to a quantitative understanding of the dissipative, irreversible evolution and loss of coherence generated during a quantum computation. Therefore, the formalism and dynamical model of quantum measurement introduced in this paper will be of fundamental importance to the engineering of future quantum devices which are able to evade wave function collapse for extended periods of time.

Finally, we would like to comment further on the relationship between our approach and the many-worlds interpretation of quantum mechanics~\cite{everett1957hugh,deutsch2011fabric}. The many-worlds interpretation of quantum mechanics states that quantum measurement is a dynamical process that also obeys the time-reversible Schrödinger equation. However, at no stage in this approach is there an attempt to rigorously understand the spectral properties or classification of the type of dynamical system produced by a quantum measurement interaction. Therefore, naive application of the Schrödinger equation leads to the unitary evolution of the macroscopic system, with decoherence somehow accounting for the formation of a mixed state. With every measurement that is made, the wavefunction is supposed to somehow transform into a particular eigenstate, accompanied by the splitting of the universe into 
a vast proliferation of separate universes that evolve henceforth independently of one another. This interpretation obviously suffers from the deficiencies of decoherence theory and makes no attempt to connect to the second law of thermodynamics. In stark contrast, our work instead demonstrates objective, probabilistic collapse of the wave function and the dynamical
origin of the Born rule by simply viewing the quantum measurement as constituting a macroscopic mixing dynamical system. We are therefore able to explain why quantum measurement is an entropy increasing process as well as the irreversible collapse of the wave function in accordance with the Born rule.

\section{Conclusions}

In summary, we have provided a complete and objective account of the emergence of the arrow of time, irreversibility, equilibrium states and measurement in quantum mechanics. This has been done without any modification of
the underlying theoretical structure of quantum mechanics. Time asymmetry emerges as the result of the thermodynamic limit of large quantum systems. As a result, an isolated system containing macroscopic numbers of particles can acquire a direction of time from the complexity of the dynamics. This allows us to reconcile the second law of thermodynamics with reversible microscopic dynamics for quantum mechanical systems. Importantly, this leads to the establishment of mixed density matrix distributions from initially pure states. By building on the measurement problem originally outlined by von Neumann~\cite{von2018mathematical} and Bohm~\cite{bohm2012quantum}, we demonstrate that the quantum mechanical measurement process directly follows from the intrinsic dynamics of the macroscopic measurement device interacting with the subsystem being measured. Our analysis shows that quantum mechanical measurement is a special case of the general dissipative dynamics of mixing quantum systems which contain continuous spectra, leading to the entropy increase of the universe. Finally, we have demonstrated how definite outcomes emerge from a given experiment as the macroscopic equilibrium states resulting from the measurement interaction become disjoint, corresponding to distinct physical realities. The objective collapse of the wave function to a given outcome is inherently probabilistic and is in exact agreement with the Born rule. The theory presented here has far-reaching ramifications, going well beyond decoherence theory in the
context of open quantum systems and opening the way \emph{inter alia} to a systematic development of quantum kinetic equations 
which govern transport phenomena and other non-equilibrium processes.


\section*{Acknowledgments}

The authors thank Michael Berry and Pragya Shukla for their insightful comments on the manuscript.  P.V.C. acknowledges funding support from the UK Engineering and Physical Sciences Research Council through UKCOMES (EP/R029598/1) and SEAVEA (EP/W007711/1). 

\appendix

\section{Algebraic quantum theory and abstract state spaces}~\label{app:aqm}

Algebraic quantum theory emerged from the identification that observables in quantum mechanics can be associated with a more general structure than that contained within a vector space. The formulation is particularly important when dealing with infinite quantum systems such as those obtained within the thermodynamic limit or quantum fields. In particular, algebraic quantum theory introduces the following postulates~\cite{davies1976quantum,fewster2020algebraic}:
\begin{enumerate}
    \item A physical system is described by a unital $*$-algebra $\mathcal{A}$, with self-adjoint elements that are the observables. 
    \item Quantum states are given by normalized linear functionals $\omega : \mathcal{A} \rightarrow \mathbbm{C}$. This means that we must have $\omega(A^*A) \geq 0$,   $\omega(\mathbbm{1})=1$ and the linear property $\omega(\alpha A+\beta B) = \alpha\omega(A)+\beta\omega(B)$. 
\end{enumerate}
A $*$-algebra $\mathcal{A}$ is simply an algebra over $\mathbbm{C}$, with a map $*: \mathcal{A}\rightarrow\mathcal{A}$ called an involution. A unital algebra requires the existence of an identity element such that  $A\mathbbm{1}=\mathbbm{1}A=A\hspace{1mm} \forall A \in \mathcal{A}$. Particularly useful $*$-algebras are Banach algebras in which the $*$-algebra can be equipped with a norm $||\cdot||$ that yields a complete algebra such that $||A^*||=||A||$. This concept naturally leads to the $C^*$-algebra which further requires that the Banach algebra has the property $||A^*A||=||A||^2$.  

The algebraic quantum state $\omega$ gives the expected value of an observable $A$, provided that $A^*=A$. Therefore, $\omega(A)$ corresponds to the expected value of $A$ if measured in the state $\omega$. In particular, if $\mathcal{A}$ is an algebra of bounded operators acting on the Hilbert space $\mathcal{H}$, then a vector $\psi \in \mathcal{H}$ induces a vector state on $\mathcal{A}$ with 
\begin{gather}
    \begin{split}
        \omega(A) = \braket{\psi|A\psi} \ .
    \end{split}
\end{gather}
Likewise, a positive trace-class operator $\rho$ on $\mathcal{H}$ can also induce a state on $\mathcal{A}$ as 
\begin{gather}
    \begin{split}
        \omega(A) = \text{tr}\{\rho A\}  \ .
    \end{split}
\end{gather}
However, it is crucial to note that the algebraic quantum state is fundamentally mathematical in nature and does not need to be defined from a specific Hilbert space representation. 

The more general axiomatic approach to defining an abstract state space, useful in the unification of probability theory and quantum theory, can be presented as follows~\cite{davies1976quantum}. An abstract state space can be defined as a real Banach space $V$ that is partially ordered by a cone $V^+$ where:
\begin{enumerate}
    \item $V^+$ is a closed set in $V$
    \item If $x,y\in V^+$, then $||x||+||y|| = ||x+y||$
    \item If $x\in V$ and $\epsilon>0, \exists\hspace{1mm} x_1,x_2\in V^+ : x=x_1-x_2$ and $||x_1||+||x_2||<||x||+\epsilon$
\end{enumerate}
In any state space the norm on $V^+$ can be extended to a positive linear functional $\tau : V \to \mathbbm{R}$:
\begin{equation*}
    |\tau(x)| \leq ||x|| \hspace{2.5mm}\forall\hspace{1mm} x\in V
\end{equation*}
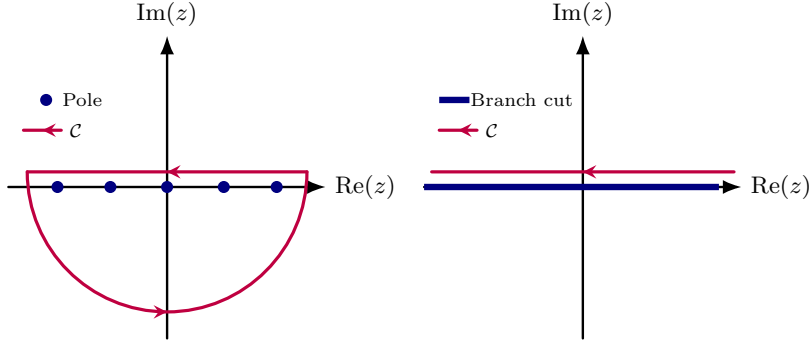
\begin{figure*}[ht]
\centering
    \begin{tikzpicture}[>=Latex, line cap=round, line join=round, font=\small]

\tikzset{
  axis/.style={->, line width=0.9pt},
  cont/.style={<->, line width=0.9pt},
  pole/.style={circle, line width=0.9pt, minimum size=4pt,
               fill=blue!50!black, draw=none, inner sep=1.6pt},
  cut/.style={line width=2.4pt, draw=blue!50!black, line cap=butt},
  poles/.style={circle, line width=0.9pt, minimum size=6pt,
               fill=purple!50!black, draw=none, inner sep=1.6pt},
  cutI/.style={line width=2.4pt, draw=purple!50!black, line cap=butt}
}

\def\L{0.5}
\def\R{5}
\def\xmin{-2.1}
\def\xmax{2.1}
\def\ymin{-2.0}
\def\ymax{2.0}

\begin{scope}[shift={(\L,0)}]
  \draw[axis] (\xmin,0) -- (\xmax,0) node[right] {$\mathrm{Re}(z)$};
  \draw[axis] (0,\ymin) -- (0,\ymax) node[above] {$\mathrm{Im}(z)$};

\def\Rcontour{1.85}
\def\yeps{0.20}

\draw[
  purple,
  very thick,
  postaction={
    decorate
  },
  decoration={
    markings,
    mark=at position 0.5 with {\arrow{stealth}}
  }
]
(\Rcontour,\yeps) -- (-\Rcontour,\yeps);

\draw[
  purple,
  very thick,
  postaction={
    decorate
  },
  decoration={
    markings,
    mark=at position 0.5 with {\arrow{stealth}}
  }
]
(-\Rcontour,\yeps)
arc[
  start angle=180,
  end angle=360,
  radius=\Rcontour
];

     \node[pole] at (0.0,0.0) {};
      \node[pole] at (0.75,0) {};
  \node[pole] at (1.45,0) {};
  \node[pole] at (-0.75,0) {};
  \node[pole] at (-1.45,0) {};


  \node[pole] at (-1.6,1.15) {};
  \node[anchor=west] at (-1.5,1.15) {\scriptsize Pole};
    \draw[
    purple,
  very thick,
  postaction={
    decorate
  },
  decoration={
    markings,
    mark=at position 0.5 with {\arrow{stealth}}
  }
](-1.4,0.75) -- (-1.9,0.75);
\node[anchor=west] at (-1.4,0.75) {\scriptsize $\mathcal{C}$};
  
\end{scope}

\begin{scope}[shift={(\R+1,0)}]
  \draw[axis] (\xmin,0) -- (\xmax,0) node[right] {$\mathrm{Re}(z)$};
  \draw[axis] (0,\ymin) -- (0,\ymax) node[above] {$\mathrm{Im}(z)$};

  \node[above left] at (1.5,1.2) {};
  \node[above left] at (1.5,0.6) {};

  \draw[cut] (\xmin,0) -- (\xmax-0.3,0);

  \def\Rcontour{2.00}
\def\yeps{0.20}

\draw[
    purple,
  very thick,
  postaction={
    decorate
  },
  decoration={
    markings,
    mark=at position 0.5 with {\arrow{stealth}}
  }
]
(\Rcontour,\yeps) -- (-\Rcontour,\yeps);

\draw[cut] (-1.90,1.15) -- (-1.5,1.15);
\node[anchor=west] at (-1.6,1.15) {\scriptsize Branch cut};
  
\draw[
    purple,
  very thick,
  postaction={
    decorate
  },
  decoration={
    markings,
    mark=at position 0.5 with {\arrow{stealth}}
  }
](-1.4,0.75) -- (-1.9,0.75);
\node[anchor=west] at (-1.4,0.75) {\scriptsize $\mathcal{C}$};

\end{scope}


\end{tikzpicture}

\caption{Different algebraic structures of the resolvent of the Liouvillian due to the nature of its spectrum. The left plot shows a finite quantum system that oscillates unitarily forever as the Laplace transform can be written in terms of a single closed contour. The right plot shows that, for non-integrable systems with an absolutely continuous spectrum, the Laplace transform cannot be written in terms of a closed contour on the first Riemann sheet. This leads to dissipative semi-group evolution. Evaluations of these integrals gives the Frobenius-Perron operator of the system: $e^{-i\mathcal{L}t}=\int_{\mathcal{C}}\frac{dz}{2\pi i}e^{-izt}(z-\mathcal{L})^{-1}$.  
}
\label{fig:inv_laplace}

\end{figure*}
and 
\begin{equation*}
    \tau(x)= ||x|| \hspace{2.5mm}\forall\hspace{1mm} x\in V^+ 
\end{equation*}
The states are then defined as the elements of $\{x\in V^+:\tau(x)=1\}$. 

For example, let $\mathcal{A}$ be a C*-algebra. $V$ can then be defined as the space of all bounded linear functionals $\omega$ on $\mathcal{A}$ which are `self-adjoint' as $\omega(A^*)=\overline{\omega(A)} \hspace{1mm}\forall\hspace{1mm} A \in \mathcal{A}$.  The ordered cone on $V$, denoted as $V^+$ is then given by 
\begin{equation*}
    V^+ := \{\omega \in V:\omega(A^*A) \geq 0 \hspace{1mm}\forall\hspace{1mm} A \in \mathcal{A}\} \ .
\end{equation*}
The norm of these states is then given by the dual space norm:
\begin{gather}
\begin{split}
    ||\omega||= \sup\{|\omega(A)|:A \in\mathcal{A}, ||A||\leq 1\} \ .
\end{split}
\end{gather}
As $\mathcal{A}$ has an identity element, $\mathbbm{1}$, we can write the norm of the state as 
\begin{gather}
    \begin{split}
        ||\omega||= \omega(\mathbbm{1}) = 1 \ .
    \end{split}
\end{gather}

Throughout this paper we will make use of the Banach dual space defined as 
\begin{gather}
    \begin{split}
        &\mathcal{A^*} = \\
        &\{\omega: \mathcal{A}\to\mathbbm{C}\hspace{0.5mm}|\hspace{0.5mm}\omega(\alpha A+\beta B) =\alpha\omega(A)+\beta\omega(B)\hspace{1mm} \forall \hspace{0.5mm} A,B \in \mathcal{A} \} .
    \end{split}
\end{gather}
The Banach dual space is therefore given by the set of all bounded, complex functionals on the Banach space, $\mathcal{A}$. The physical state space is then defined as 
\begin{gather}
    \begin{split}
        \mathcal{S}(\mathcal{A}) = \{\omega \in \mathcal{A^*}: \omega \geq 0, \omega(\mathbbm{1})=1\} \ .
    \end{split}
\end{gather}
It is clear that $\mathcal{S}(\mathcal{A})\subset \mathcal{A^*}$. By introduction of the Banach dual space, we can define the bilinear dual pairing $\langle\cdot,\cdot \rangle:\mathcal{A}\times \mathcal{A^*}\to\mathbbm{C}$ such that 
\begin{gather}
    \begin{split}
        \langle A,\omega \rangle = \omega(A) \ .
    \end{split}
\end{gather}
Clearly for expectation values, we must have $\langle A^*,\omega \rangle = \overline{\langle A,\omega \rangle }$. We can now choose to analyze the dynamics on the Banach space or the dual Banach space via the relation: $\langle A_t,\omega \rangle = \langle A,\omega_t \rangle $. The time-dependent equations of motion are given by 
\begin{subequations}
    \begin{align}
        \begin{split}
            \frac{\partial}{\partial t} \omega_{t} &= -i\mathcal{L}\omega_{t}
        \end{split}\\
        \begin{split}
            \frac{\partial}{\partial t} A_{t} &= i\mathcal{L^*}A_{t} \ ,
        \end{split}
    \end{align}
\end{subequations}
where $\mathcal{L}:\mathcal{A^*}\to\mathcal{A^*}$ and the dual generator on the state of observables is $\mathcal{L^*}:\mathcal{A}\to\mathcal{A}$. Using the dual pairing, it is straightforward to demonstrate 
\begin{gather}
    \begin{split}
        \langle A,\mathcal{L}\omega_t \rangle = -  \langle \mathcal{L^*}A_t,\omega_0 \rangle \ .
    \end{split}
\end{gather}
This equation is equivalent to that given in the main text: $[\mathcal{L}\omega_{t}](A)=-\omega_{0}(\mathcal{L^*}A_t)$. However, in the main text we denote $\mathcal{L^*}$ with the same symbol $\mathcal{L}$ for notational simplicity. In the thermodynamic limit, where the spectrum of the generator $\mathcal{L}$ is absolutely continuous, we are particularly interested in the case where the Banach algebra is given by a type III$_{1}$ von Neumann algebra~\cite{bratteli2012operator}. This is because the state $\omega$ can no longer be expressed as a trace-class operator~\cite{witten2018aps,witten2022does}. In fact, the singular thermodynamic limit requires the transition from type I to type III von Neumann algebras~\cite{witten2018aps,witten2022does}.

\section{Algebraic structure of the resolvent and reversible unitary evolution}~\label{app:1}

For finite isolated quantum systems, the spectrum of the Liouvillian is discrete. Therefore the eigenstates exist in the Banach space and the resolvent of the Liouvillian is a simple meromorphic function containing discrete poles at the energy differences of the system. The eigenvalue equation for the generator and its dual is given by 
\begin{subequations}
    \begin{align}
         \begin{split}
        \mathcal{L}\omega_{\lambda} &= \varepsilon_{\lambda} \omega_{\lambda} \ , \hspace{2mm} \omega_{\lambda} \in \mathcal{A^*}
    \end{split}\\
    \begin{split}
        \mathcal{L^*}A_{\lambda} &= -\varepsilon_{\lambda} A_{\lambda} \ , \hspace{2mm} A_{\lambda} \in \mathcal{A} \ .
    \end{split}
    \end{align}
\end{subequations}
It is straightforward to demonstrate that $\langle A_{\lambda},\omega_{\lambda'}\rangle=\omega_{\lambda'}(A_{\lambda})=\delta_{\lambda\lambda'}$. 
Using these abstract eigenstates, we can write the resolvent acting on an element of the dual algebra $\mathcal{A^*}$ as 
\begin{gather}
    \begin{split}
        (z-\mathcal{L})^{-1} &= \sum_{\lambda} \frac{\omega_{\lambda}\langle A_{\lambda},\cdot\rangle}{z-\varepsilon_{\lambda}}    \ .
    \end{split}
\end{gather}
Taking the inverse Laplace transform of this expression for $t>0$ gives 
\begin{gather}
    \begin{split}
        \sum_{\lambda}i\int^{\infty+i\gamma}_{-\infty+i\gamma}\frac{dz}{2\pi}\ e^{-izt} \frac{\omega_{\lambda}\langle A_{\lambda},\cdot\rangle}{z-\varepsilon_{\lambda}} &=  \sum_{\lambda} e^{-i\varepsilon_{\lambda}t}\omega_{\lambda}\langle A_{\lambda},\cdot\rangle\\
        &=e^{-i\mathcal{L}t} \ .
    \end{split}
\end{gather}
This expression is valid $\forall\hspace{1mm} t \in \mathbbm{R}$. This is because the simple meromorphic structure means that the result for both $t\gtrless0$ can be obtained from a single integration over a single closed contour. Therefore, the unitary group is recovered and reversible evolution results. For the case where the spectrum of $\mathcal{L}$ is absolutely continuous, the resolvent exhibits a branch cut across the entire real axis and so the time-evolution operator cannot be written in terms of a single closed contour (see Fig.~\ref{fig:inv_laplace}).

\section{Continuous spectrum in algebraic quantum theory}~\label{app:spec_cont}

We are particularly interested in the case where the spectrum of the Liouvillian is absolutely continuous. The generator $\mathcal{L}$ acts on the dual Banach space. The eigenvalue problem is the following 
\begin{gather}
    \begin{split}
        \mathcal{L}\Omega_{\lambda}= \lambda \Omega_{\lambda} \ .
    \end{split}
\end{gather}
Due to the absolutely continuous nature of the spectrum $\sigma(\mathcal{L})$, the eigenstates $\Omega_{\lambda}$ do not exist in the usual dual Banach space $\mathcal{A^*}$. This is because the norm $||\Omega_{\lambda}||_{\mathcal{A^*}}$ is not bounded. As they cannot be normalized under the standard Banach space norm topology, they cannot reside in $\mathcal{A^*}$. Likewise, the absolutely continuous spectrum can be expressed with respect to the Banach space generator:
\begin{gather}
    \begin{split}
        \mathcal{L^*}A_{\lambda} = -\lambda A_{\lambda} \ .
    \end{split}
\end{gather}
Similarly, the eigenstates $A_{\lambda}$ cannot exist in the Banach space $\mathcal{A}$ as they are not bounded by the norm topology. To rigorously construct the eigenstates and the space they exist in, we need to construct a Gel'fand triple~\cite{gel2014generalized}. This consists of an appropriate test observable space $\Phi \subset \mathcal{A}$ defined such that 
\begin{enumerate}
    \item $\Phi$ is dense in $\mathcal{A}$.
    \item $\Phi\to\mathcal{A}$ is continuous. 
    \item $\mathcal{L^*}\Phi\subseteq \Phi$.
    \item $\mathcal{L^*}:\Phi\to\Phi$ is continuous in the finer topology of $\Phi$. 
\end{enumerate}
As the topology of $\Phi$ is stronger than that of the C*-norm topology, any bounded functional restricts continuously to $\Phi$. Using $\Phi$, we can immediately define the continuous dual $\Phi'$. Therefore, we have the following triple: $\mathcal{S}(\mathcal{A})\subset \mathcal{A^*}\subset \Phi'$. From the continuous dual space $\Phi'$, we can define the generalized state generator $\mathcal{L}^{\text{x}}:\Phi'\to\Phi'$ such that 
\begin{gather}
    \begin{split}
        \langle A , \mathcal{L}^{\text{x}}\Omega \rangle = -\langle \mathcal{L^*}A , \Omega \rangle \ , \hspace{2mm} \forall \hspace{1mm} A\in\Phi \ .
    \end{split}
\end{gather}
We can then write the generalized eigenfunctional in terms of the extended generator as 
\begin{gather}
    \begin{split}
        \mathcal{L}^{\text{x}}\Omega_{\lambda} = \lambda \Omega_{\lambda} \ , \hspace{2mm}\Omega_{\lambda} \in \Phi' \ .
    \end{split}
\end{gather}
with 
\begin{gather}
    \begin{split}
        \langle \mathcal{L^*}A , \Omega_{\lambda} \rangle = - \langle A , \mathcal{L}^{\text{x}}\Omega_{\lambda} \rangle = -\lambda \langle A , \Omega_{\lambda}\rangle \ , \hspace{2mm} \forall\hspace{1mm} A \in \Phi \ .
    \end{split}
\end{gather}
In terms of the functional notation, we have 
\begin{gather}
    \begin{split}
        \Omega_{\lambda}(\mathcal{L}^*A) = -\lambda\Omega_{\lambda}(A) \ .
    \end{split}
\end{gather}
We can now write the generalized eigenvalue equations as
\begin{subequations}~\label{eqs:cont_liou}
    \begin{align}
        \begin{split}
            \mathcal{L}^{\text{x}}\Omega_{\lambda} &= \lambda \Omega_{\lambda} \ 
        \end{split}\\
        \begin{split}
            (\mathcal{L^*})^{\text{x}}A_{\lambda} &= -\lambda A_{\lambda} \ ,
        \end{split}
    \end{align}
\end{subequations}
with the relationship $\langle A_{\lambda},\Omega_{\lambda'}\rangle=\Omega_{\lambda'}(A_{\lambda})=\delta(\lambda-\lambda')$ as a distributional pairing. Using these states, we can write the Green's function as 
\begin{gather}
    \begin{split}
        G_{A}(z) &= \langle A , (z-\mathcal{L})^{-1}\omega_0\rangle \\
        &= \int^{\infty}_{-\infty} d\lambda\ \frac{\langle A , \Omega_{\lambda}\rangle \langle A_{\lambda} , \omega_{0}\rangle}{z-\lambda} \\ 
        &= \int^{\infty}_{-\infty} d\lambda\ \frac{ \Omega_{\lambda}(A) c_{\lambda}(\omega_0)}{z-\lambda} \ .
    \end{split}
\end{gather}
We have used the notation $c_{\lambda}(\omega_0)=\langle A_{\lambda} , \omega_{0}\rangle$ to emphasize the extension of the bilinear dual pairing to include the continuous eigenstates. 
To summarize, for the continuous spectrum, we have the triplet: $\mathcal{S}(\mathcal{A})\subset \mathcal{A^*}\subset \Phi'$ such that 
\begin{itemize}
    \item $\omega_{t}\in\mathcal{S}(\mathcal{A})$ which is a physical state. 
    \item $\mathcal{L}$ acts on an appropriate subspace of $\mathcal{A^*}$.
    \item $\Omega_{\lambda}\in\Phi'$ are generalized continuous-spectrum modes. 
\end{itemize}
Therefore, it is straightforward to realise that for systems that possess an absolutely continuous spectrum of the Liouvillian, the generalized eigenstates $\Omega_{\lambda}$ do not satisfy the properties required in order to be referred to as physical states in the quantum theory.

\section{Analytic continuation and complex eigenvalues}~\label{app:continuation}

Due to the continuous spectrum of $\mathcal{L}$, the resolvent $(z-\mathcal{L})^{-1}$ and therefore the Green's function $G_{A}(z)$ possess a branch cut across the entire real axis. This is a result of the discontinuity:
\begin{gather}
    \begin{split}
        G_{A}(\omega-i0^+)-G_{A}(\omega+i0^+) = 2\pi i \Gamma_{A}(\omega) \ ,
    \end{split}
\end{gather}
where $\Gamma_{A}(\omega) = \langle A , \delta(\omega-\mathcal{L})\omega_0\rangle$. Remembering our convention, we have $\langle A , (z-\mathcal{L})^{-1}\omega_0\rangle=\langle (z+\mathcal{L^*})^{-1}A , \omega_0\rangle$. For a mixing system, it is straightforward to realize that for $A \in \Phi$ and $\omega_0\in \mathcal{S}(\mathcal{A})$, $G_{A}(z)$ admits a meromorphic continuation through the branch cut: $G^{\text{II}}_{A}(z)$. The continuation requires the seam condition to be obeyed: $G^{\text{II}}_{A}(\omega-i0^+)=G_{A}(\omega+i0^+)$. 

Suppose that the continuation reveals a series of simple poles at $z=z_J$ that must lie beneath the real axis $\text{Im}z_{J} \leq 0$ due to causality~\cite{coveney2026thesis}. Therefore, we can write 
\begin{gather}
    \begin{split}
        G^{\text{II}}_{A}(z) = \sum_{J}\frac{\tilde{\omega}_{J}(A)}{z-z_J} + G^{\text{II,c}}_{A}(z) \ ,
    \end{split}
\end{gather}
where $G^{\text{II,c}}_{A}(z)$ is the background continuum contribution from the second Riemann sheet~\cite{gaspard2005chaos,bohm2013quantum,reed1979iii}. 
It is therefore clear that $\tilde{\omega}_{J}(A)=\text{Res}_{z=z_J}G^{\text{II}}_{A}(z)$ and must consist of a bilinear form on $\Phi\times S(\mathcal{A})$. Therefore, we can factorize the residue as 
\begin{gather}
    \begin{split}
        \tilde{\omega}_{J}(A) &= \langle A,\tilde{\Omega}_{J}\rangle \langle \tilde{A}_J,\omega_0\rangle \\
        &= \tilde{\Omega}_J(A)\tilde{c}_J(\omega_0) \ .
    \end{split}
\end{gather}
It is clear that neither one of the states $\tilde{\Omega}_{J}$ or $\tilde{A}_J$ exist in $\mathcal{A^*}$ or $\mathcal{A}$. Through this factorization, we can infer that the residues must satisfy the equation
\begin{subequations}~\label{eq:ext_liouville}
    \begin{align}
    \begin{split}
        \mathcal{L}^{\text{x}}_{+}\tilde{\Omega}_J &= z_{J}\tilde{\Omega}_J 
    \end{split}\\
    \begin{split}
        (\mathcal{L^*})^{\text{x}}_{+}\tilde{A}_J &= -z_{J}\tilde{A}_J 
    \end{split}
\end{align}
\end{subequations}
where $\mathcal{L}^{\text{x}}_{+}$ represents the extension of $\mathcal{L}^{\text{x}}$ to the generalized dual space $\Phi'_{+}$. The resonance states $\tilde{\Omega}_{J}$ with eigenvalue $z_{J}\neq 0$ are not physical states as $\tilde{\Omega}_{J}(\mathbbm{1}) = 0 $. The only state that is physical is the zero resonance state $\langle\mathbbm{1},\mathcal{L}^{\text{x}}_{+}\tilde{\Omega}_0\rangle = 0$, which gives $\langle\mathbbm{1},\tilde{\Omega}_0\rangle=\tilde{\Omega}_0(\mathbbm{1}) = 1$. Therefore, the equilibrium state $\tilde{\omega}_0(A) = \langle A,\tilde{\Omega}_{0}\rangle \langle \mathbbm{1},\omega_0\rangle = \tilde{\Omega}_0(A)$. It is important to note that the equilibrium state can be degenerate in general, depending on the number of conserved quantities. This state is the invariant measure of the dynamics. 

After contour deformation, on the second Riemann sheet we have 
\begin{gather}
    \begin{split}
        \omega_t(A) &= \oint_{\mathcal{C}}\frac{dz}{2\pi i} e^{-izt}  G^{\text{II}}_{A}(z) \\
        &= \sum_{J}e^{-iz_{J}t} \tilde{\omega}_{J}(A) + G^{\text{II,c}}_{A}(t) \\
        &= \langle A,\tilde{\Omega}_{0}\rangle \langle \mathbbm{1},\omega_0\rangle+\sum_{J\neq 0}e^{-iz_{J}t} \langle A,\tilde{\Omega}_{J}\rangle \langle \tilde{A}_J,\omega_0\rangle \\
        &+ G^{\text{II,c}}_{A}(t)
    \end{split}
\end{gather}
for $t>0$, where $\text{Im}z_{J}\leq 0$. In the asymptotic limit, we find that $\lim_{t\to\infty}\omega_t(A)=\tilde{\omega}_0(A)$, which is the invariant measure. The physical state $\omega_t(A)$ remains in $\mathcal{S}(\mathcal{A})$ despite the fact that the resonance functionals $\tilde{\Omega}_J$ with eigenvalue $z_J\neq 0$ are not positive. This is because the positivity is recovered by the complete sum of the pole and background terms. The equilibrium state $\tilde{\omega}_0=\tilde{\Omega}_0$ is simultaneously a generalized eigenfunctional of $\mathcal{L}^{\text{x}}_{+}$ and a genuine physical state. Importantly, as the spectrum of the Liouvillian is symmetric (see Eq.~\ref{eq:spec_liouville}), the spectrum of $\mathcal{L}^{\text{x}}_{+}$ is symmetric such that the eigenvalues appear in pairs: $ z_{J} = \pm\epsilon_{J} - i\gamma_{J}$. This symmetry is depicted in Fig.~\ref{fig:causal_sheets}(b).

We can include the set of  pole states and continuous energies generated by the branch cuts in the complex plane on the second Riemann sheet to express the background continuous contributions also in terms of explicit eigenstates of the continued Liouvillian as:
\begin{subequations}
    \begin{align}
        \begin{split}
            \mathcal{L}^{\text{x}}_{+}\tilde{\Omega}_J &= z_{J}\tilde{\Omega}_J 
        \end{split}\\
        \begin{split}
            \mathcal{L}^{\text{x}}_{+}\tilde{\Omega}_\xi &= \xi\tilde{\Omega}_\xi 
        \end{split}
    \end{align}
\end{subequations}
where $\text{Im}z_{J}\leq 0$ and $\text{Im}\xi< 0$. Likewise, it is also clear that the continuum solutions appear in pairs: $\xi = \pm\text{Re}\xi+i\text{Im}\xi$. We also have the extended eigenstates for the dual generator as 
\begin{subequations}
    \begin{align}
        \begin{split}
            (\mathcal{L^*})^{\text{x}}_{+}\tilde{A}_J &= -z_{J}\tilde{A}_J 
        \end{split}\\
        \begin{split}
            (\mathcal{L^*})^{\text{x}}_{+}\tilde{A}_\xi &= -\xi\tilde{A}_\xi 
        \end{split}
    \end{align}
\end{subequations}
such that $\langle \tilde{A}_J,\tilde{\Omega}_{J'}\rangle = \tilde{\Omega}_{J'}(\tilde{A}_J)=\delta_{JJ'}$ and $\langle \tilde{A}_\xi,\tilde{\Omega}_{\xi'}\rangle = \tilde{\Omega}_{\xi'}(\tilde{A}_\xi)=\delta(\xi-\xi')$. Using these generalized states we can write 
\begin{gather}
    \begin{split}
        \omega_t(A) 
        &= \sum_{J}e^{-iz_{J}t} \langle A,\tilde{\Omega}_{J}\rangle \langle \tilde{A}_J,\omega_0\rangle \\
        &+ \int_{\Gamma_{\text{II}}} d\xi e^{-i\xi t}  \langle A,\tilde{\Omega}_{\xi}\rangle \langle \tilde{A}_\xi,\omega_0\rangle \ .
    \end{split}
\end{gather}
Here, the notation $\Gamma_{\text{II}}$ is general as we can always orient the complex continuum branch cuts on the second Riemann sheet to run parallel to the imaginary axis~\cite{gaspard2005chaos}. 
Importantly, $\mathcal{L}^{\text{x}}_{+}$ gives rise to a forward-time semi-group generator on the generalized dual space $\Phi'_{+}$. This means that we can define $\hat{U}^{\text{x}}_{+}(t) = \theta(t)e^{-i\mathcal{L}^\text{x}_{+}t}$, such that $\hat{U}^{\text{x}}_{+}(t)\hat{U}^{\text{x}}_{+}(s)=\hat{U}^{\text{x}}_{+}(t+s)$ for $t,s >0$ only. 

Due to the time symmetry of the equations, we can also construct an advanced continuation, $\mathcal{L}^{\text{x}}_{-}$, that gives rise to a backward-time semi-group generator on the generalized dual space $\Phi'_{-}$. However, this would involve equilibrium being obtained in the past and is not compatible with physical reality. This arises due to the relation $\hat{U}_{-}(t) = [\hat{U}_{+}(-t)]^\dag$, which leads to $\mathcal{L}^{\text{x}}_{-}=(\mathcal{L}^{\text{x}}_{+})^\dag$ on the extended Liouville-Hilbert space. Likewise, using the analytic continuation from the advanced direction we obtain a backward-time semi-group where   $\hat{U}^{\text{x}}_{-}(t) = \theta(-t)e^{-i\mathcal{L}^\text{x}_{-}t}$, such that $\hat{U}^{\text{x}}_{-}(t)\hat{U}^{\text{x}}_{-}(s)=\hat{U}^{\text{x}}_{-}(t+s)$ for $t,s <0$ only. However, this generator is nonphysical. 

\section{Emergence of the time operator in quantum mechanics}~\label{app:prigogine}

Using the continuous eigenstates of the Liouvillian given in Eqs~\ref{eqs:cont_liou}, it is possible to construct the time operator in quantum mechanics. By defining the Fourier transforms, using Stone's theorem, we can write 
\begin{subequations}
    \begin{align}
    \begin{split}
        \Gamma_t &= \int^{\infty}_{-\infty} \frac{d\lambda}{2\pi}e^{-i\lambda t}\Omega_{\lambda} 
    \end{split}\\
    \begin{split}
        A_t &= \int^{\infty}_{-\infty} \frac{d\lambda}{2\pi}e^{i\lambda t}A_{\lambda} \ .
    \end{split}
\end{align}
\end{subequations}
The time operator, $\mathcal{T}:\mathcal{A}^*\rightarrow\mathcal{A}^*$, can then be defined as 
\begin{gather}
    \begin{split}~\label{eq:time_operator}
        \mathcal{T} = \int^{\infty}_{-\infty} dt\ t \Gamma_t \langle A_t,\cdot\rangle \ ,
    \end{split}
\end{gather}
such that $[\mathcal{L},\mathcal{T}] = i$. For $\mathcal{T}$ to exist requires $\lambda\in(-\infty,\infty)$ such that the spectral measure of $\mathcal{L}$ is absolutely continuous. This is precisely the case for many quantum systems in the thermodynamic limit. The time-operator was first introduced by Misra, Prigogine and Courbage in the context of classical dynamical systems~\cite{misra1979deterministic,prigogine1982being}. The time operator defined by Eq.~\ref{eq:time_operator} provides an objective measure of the age of a macroscopic quantum system in the thermodynamic limit.

\section{Lyapunov functional and Entropy}~\label{app:lyapunov}

It is clear that the dynamics of a non-integrable mixing system is governed by the continued Liouvillian, $\mathcal{L}^{\text{x}}_{+}$. Using $\mathcal{L}^{\text{x}}_{+}$, we can generate a Lyapunov functional by writing the equation-of-motion of the state 
\begin{gather}
    \begin{split}~\label{eq:markov}
        \omega_{t} = e^{-i\mathcal{L}^{\text{x}}_{+}t}\omega_0 \ .
    \end{split}
\end{gather}
Resolving the identity, we have 
\begin{gather}
    \begin{split}
        \omega_{t} = \sum_{J}e^{-iz_Jt}\langle\tilde{A}_J,\omega_0\rangle \tilde{\Omega}_{J} + \int_{\Gamma_{\text{II}}} d\xi e^{-i\xi t}   \langle \tilde{A}_\xi,\omega_0\rangle \tilde{\Omega}_{\xi} \ .
    \end{split}
\end{gather}
It is straightforward to show that $\langle\mathbbm{1},\omega_t\rangle=\omega_t(\mathbbm{1})=1$, as stated in the main text and required in order for $\omega_t$ to be a quantum state. Therefore, $\omega_t \in \mathcal{A}^*$. 
As the eigenvalues are complex, the generalization of an associated state in the biorthogonal theory is constructed as~\cite{brody2013biorthogonal,coveney2024cc_se}
\begin{gather}
    \begin{split}
        \tilde{\omega}_{t} = \sum_{J}e^{iz^*_Jt}\overline{\langle\tilde{A}_J,\omega_0\rangle} \tilde{A}_{J} + \int_{\Gamma_{\text{II}}} d\xi e^{i\xi^* t}   \overline{\langle \tilde{A}_\xi,\omega_0\rangle} \tilde{A}_{\xi} \ .
    \end{split}
\end{gather}
Clearly $\tilde{\omega}_{t}$ is constructed from the eigenstates of the dual Liouvillian (see Eqs~\ref{eq:ext_liouville}). As it is the state associated to $\omega_t$, it is therefore an element of the Banach space such that: $\tilde{\omega}_{t}\in\mathcal{A}$. 
Taking the dual pairing, we immediately obtain the Lyapunov functional as 
\begin{gather}
    \begin{split}
        H_t &= \langle\tilde{\omega}_{t},\omega_{t}\rangle - |\langle\tilde{A}_0,\omega_0\rangle|^2 \\
        &=\sum_{J\neq0}|\langle\tilde{A}_J,\omega_0\rangle|^2 e^{2\text{Im}z_J t } + \int_{\Gamma_{\text{II}}} d\xi   |\langle \tilde{A}_\xi,\omega_0\rangle|^2 e^{2\text{Im}\xi t} \ ,
    \end{split}
\end{gather}
remembering that $\text{Im}z_{J} < 0$ and $\text{Im}\xi <0$. We have subtracted the contribution $|\langle\tilde{A}_0,\omega_0\rangle|^2$ from invariant measure, which in general can be degenerate depending on the number of conserved quantities. We have $H_t\geq 0$, with the equality holding only at equilibrium. Differentiating this expression we also find  
\begin{gather}
    \begin{split}
        \frac{d H_t}{dt} &= 2\Bigg(\sum_{J\neq0}\text{Im}z_{J}|\langle\tilde{A}_J,\omega_0\rangle|^2 e^{2\text{Im}z_{J}t}\\
        &+\int_{\Gamma_{\text{II}}} d\xi\   \text{Im}\xi |\langle \tilde{A}_\xi,\omega_0\rangle|^2 e^{2\text{Im}\xi t} \Bigg) \\
        &\leq 0 , 
    \end{split}
\end{gather}
which strictly decreases as time increases. The equality holds only at equilibrium. The establishment of a Lyapunov functional means that we can immediately identify the second law of thermodynamics through the monotonic increase in the entropy of the isolated system. Hence, we have shown that quantum dynamics and the second law of thermodynamics are entirely reconcilable.  

In fact, we can define the relative quantity:
\begin{gather}
    \begin{split}~\label{eq:entropy_lyapunov}
        D_t &= 
        \langle\mathbbm{1},\omega_{t}(\log\omega_{t}-\log\omega_{\text{eq}})\rangle\ \\
        &= [\omega_{t}(\log\omega_{t}-\log\omega_{\text{eq}})](\mathbbm{1}),
    \end{split}
\end{gather}
remembering that $\omega_{\text{eq}}=\tilde{\omega}_{0}$ is the equilibrium state with eigenvalue $z_{0}=0$. Eq.~\ref{eq:entropy_lyapunov} is identical to the Uhlmann-Araki measure~\cite{araki1976relative,uhlmann1977relative}. By the exact semi-group Markovian time evolution of the state given in Eq.~\ref{eq:markov}, it is straightforward to see that $D_{t+s} \leq D_{s}$. Therefore, we immediately find that 
\begin{gather}
    \begin{split}
        \frac{dD_t}{dt} \leq 0 \ ,
    \end{split}
\end{gather}
with the equality holding at equilibrium. This means that Eq.~\ref{eq:entropy_lyapunov} is also a Lyapunov functional. Importantly, our analysis has revealed that the Markovian semi-group that defines the evolution of the thermodynamic system (Eq.~\ref{eq:markov}) ensures that the Uhlmann-Araki measure is a globally defined Lyapunov functional. 
Using $D_t$, we obtain the thermodynamic relative entropy as
\begin{gather}
    \begin{split}
        S_t = - k_{B}D_t \ . 
    \end{split}
\end{gather}
Therefore, we have the generalized form of Boltzmann's H-theorem for the quantum state of the isolated system:
\begin{gather}
    \begin{split}
        \frac{dS_t}{dt}  = -k_B\frac{dD_t}{dt} \geq 0 \ ,  
    \end{split}
\end{gather}
thereby recovering the second law of thermodynamics. 

\section{General equation for the reduced density matrix}~\label{app:gen_dm}

Here, we derive the exact equation-of-motion for the one-particle reduced density matrix of the subsystem $\rho^{s}_{ij}(t) = \braket{P_{ij}(t)}$ using the superoperator formalism. From the LvN equation, the equation-of-motion for the reduced density matrix can be written as 
\begin{gather}
    \begin{split}~\label{eq:eom_memory_rdm}
        &\sum_{kl}\left(i\frac{\partial}{\partial t}\delta_{ik}\delta_{jl}-h_{ij,kl}\right)\rho^{s}_{kl}(t) \\
        &= 
        \sum_{kl}\int^{t}_{0} d\Bar{t}\  \Pi_{ij,kl}(t-\Bar{t})\rho_{kl}(\Bar{t}) \ ,
    \end{split}
\end{gather}
where $h_{ij,kl}=h_{ik}\delta_{jl}-h_{jl}\delta_{ik}$ is the anti-symmetrized one-body term of the Hamiltonian in superoperator form and $\Pi_{ij,kl}$ is the Liouville self-energy. For simplicity, we assume that the initial state factorizes. This equation is of the same structure as that observed in the non-equilibrium theory formulated with respect to the Keldysh contour~\cite{schwinger1961brownian,keldysh2024diagram,stefanucci2013nonequilibrium}. Upon Laplace transformation, we have 
\begin{gather}
    \begin{split}
        \sum_{kl}\left(z\delta_{ik}\delta_{jl}-h_{ij,kl}-\Pi_{ij,kl}(z)\right) \tilde{\rho}^{s}_{kl}(z) = \rho^{s,0}_{kl} ,
    \end{split}
\end{gather}
where $\rho^{s,0}_{ij}$ is the initial reduced density matrix.
Therefore, we can write  
\begin{gather}
    \begin{split}
        \tilde{\rho}^{s}_{ij}(z) = \sum_{kl} \Big(z\mathbbm{1}-\mathbf{h}-\mathbf{\Pi}(z)\Big)^{-1}_{ij,kl} \rho^{s,0}_{kl} , 
    \end{split}
\end{gather}
from which we can define the Green's function
\begin{gather}
    \begin{split}
        G_{ij,kl}(z) = \Big(z\mathbbm{1}-\mathbf{h}-\mathbf{\Pi}(z)\Big)^{-1}_{ij,kl} \ , 
    \end{split}
\end{gather}
such that 
\begin{gather}
    \begin{split}
        \tilde{\rho}^{s}_{ij}(z) = \sum_{kl} G_{ij,kl}(z) \rho^{s,0}_{kl} \ .
    \end{split}
\end{gather}
Therefore, the dynamics of the density matrix is determined entirely by the Green's function as 
\begin{gather}
    \begin{split}~\label{eq:rdm_eom}
        \rho^{s}_{ij}(t) = \sum_{kl}G_{ij,kl}(t)\rho^{s,0}_{kl} \ ,
    \end{split}
\end{gather}
where the time domain Green's function is given by the inverse Laplace transform
\begin{gather}
    \begin{split}
        G_{ij,kl}(t)=\oint_{\mathcal{C}}\frac{dz}{2\pi i}e^{-izt} G_{ij,kl}(z) \ .
    \end{split}
\end{gather}
If the full resolvent $(z-\mathcal{L})^{-1}$ contains a branch cut due to the underlying continuous spectrum of the full many-body Hamiltonian, this structure will be `propagated' down to the reduced density matrix (which is a projection of the full resolvent). This will be clear through the emergence of the non-zero imaginary part of the self-energy: $\text{Im}\mathbf{\Pi}(\omega)\neq 0$. Causality requires that $\text{Im}(\mathbf{v}^\dag\text{Im}\mathbf{\Pi}(\omega+i0^+)\mathbf{v})\leq 0$, where $\mathbf{v}$ is an arbitrary vector of the Liouville space, such that all decay rates are positive~\cite{coveney2026thesis}. In this case, the time evolution of the system is irreversible and dissipative with the reduced density matrix retaining its hermitian structure: $\rho_{ij}(t) = \rho_{ji}^*(t)$. It is important to state that the analytically continued representation given in Eq.~\ref{eq:rdm_eom} maintains the completely positive-trace preserving (CPTP) requirements necessary in order for $\rho_{ij}(t)$ to provide a faithful representation of the quantum state as a quantum channel~\cite{park2024quasi,huang2026coupled}. 

For mixing quantum systems in the thermodynamic limit, we have demonstrated that the time evolution of the quantum state undergoes a symmetry breaking into retarded and advanced sectors. The retarded sector evolves to an equilibrium state in the future and the advanced sector evolves to an equilibrium state in the past. The boundary condition of the second law of thermodynamics means that the retarded sector represents the physical solution. In Appendix~\ref{app:lyapunov}, we explicitly demonstrated that the retarded sector evolution of the isolated system gives rise to the second law. The second law dictates the \emph{monotonic increase} in the entropy of an isolated system until it reaches its maximum value at equilibrium. 

It is important to discuss how the monotonic global increase in the entropy of the isolated system demonstrated in Appendix~\ref{app:lyapunov} relates to the increase in the entropy of the reduced density matrix whose equation of motion is given by Eq.~\ref{eq:eom_memory_rdm}. As is well known, the equation of motion for the reduced density matrix (Eq.~\ref{eq:eom_memory_rdm}) is non-Markovian as the time-dependent Liouville self-energy, $\mathbf{\Pi}(t)$, incorporates `memory' effects in the evolution. This is the result of the complex correlations in the time evolution of the subsystem that are generated due to the coupling to the `surroundings'. As a result of the structure of the dynamical self-energy, the decrease in the subsystem equivalent of Eq.~\ref{eq:entropy_lyapunov},
\begin{gather}
    \begin{split}
        D_s(t)= \text{tr}\{\rho_{s}(t)(\log\rho_s(t)-\log\rho^{\text{eq}}_{s})\} \ ,
    \end{split}
\end{gather}
is no longer monotonic: $\frac{d D_s(t)}{dt}\nless 0 $. This means that the entropy of the subsystem can oscillate due to interaction with the surroundings. However, despite these oscillations, the entropy of the subsystem will increase as a consequence of the global dissipative evolution of the isolated system. Therefore, we find that the entropy of the isolated system increases monotonically, while the entropy of the subsystem can increase non-monotonically. Importantly, the non-monotonic increase in the entropy of the subsystem is perfectly compatible with the monotonic increase in the entropy of the isolated system. This is because the second law states that it is the entropy of the \emph{isolated system} that must increase monotonically.

This represents a profoundly different physical picture to that presented by decoherence theory, which attempts to preserve the unitary evolution of the isolated system in the thermodynamic limit. As there is no such concept as the second law of thermodynamics in decoherence theory, it necessarily requires the resulting entropy change of the surroundings to always compensate any changes in the entropy of the subsystem such that the total entropy change is always zero. As our findings show, the picture implied by quantum decoherence theory is clearly not consistent. This is because the monotonic increase in the entropy of the isolated system, the non-monotonic increase in the entropy of the subsystem as well as the intrinsic dissipative dynamics of mixing quantum systems in the thermodynamic limit can all be accounted for without modification of the quantum theory.

\section{Competing dynamics and emergence of the microcanonical ensemble}

Instead of the measurement Hamiltonian in Eq.~\ref{eq:measure_ham}, if we consider the Hamiltonian 
\begin{gather}
    \begin{split}
        H &= \sum_{\substack{i\neq j \\i,j=\{+,-\}}}\Delta\epsilon a^\dag_{i}a_{j}+\int_{\mathbbm{R}^3} d\mathbf{k}\ \omega(\mathbf{k}) b^\dag_\mathbf{k}b_\mathbf{k} \\
        &+ \sum_{i=\{+,-\}}\int_{\mathbbm{R}^3} d\mathbf{k}\ \tilde{g}_{i}(\mathbf{k}) A_\mathbf{k} a^\dag_{i}a_{i} \\
        &+\frac{1}{2}\int_{\mathbbm{R}^3} d\mathbf{k}d\mathbf{q} d\mathbf{p}\ v(\mathbf{p})b^\dag_{\mathbf{k-p}}b^\dag_{\mathbf{q+}\mathbf{p}}b_{\mathbf{q}}b_{\mathbf{k}}\ ,
    \end{split}
\end{gather}
we now have a system whereby the projections of the $\pm$-spins are not conserved. This situation represents the competition between the spin precession in the external z-direction magnetic field and with the measurement of the x-axis component (for example). As a result, we arrive at the following equation-of-motion for the reduced density matrix (using $\rho_{+-}(t) = \rho^*_{-+}(t)$)
\begin{subequations}
    \begin{align}
        \begin{split}
            i\frac{\partial}{\partial t}\rho_{++}(t) &= -2i\Delta\epsilon\text{Im}\rho_{+-}(t)\\
            &+ \sum_{kl}\int^{t}_{0}dt'\ \Pi_{++,kl}(t-t')\rho_{kl}(t') \ . 
        \end{split}\\
                \begin{split}
            i\frac{\partial}{\partial t}\rho_{--}(t) &= 2i\Delta\epsilon\text{Im}\rho_{+-}(t)\\
            &+ \sum_{kl}\int^{t}_{0}dt'\ \Pi_{--,kl}(t-t')\rho_{kl}(t') \ . 
        \end{split}\\
                \begin{split}
            i\frac{\partial}{\partial t}\rho_{+-}(t) &=\Delta\epsilon\left(\rho_{--}(t) -\rho_{++}(t)\right) \\
            &+ \sum_{kl}\int^{t}_{0}dt'\ \Pi_{+-,kl}(t-t')\rho_{kl}(t') \ . 
        \end{split}
    \end{align}
\end{subequations}
Due to the thermodynamic limit, the off-diagonal decoherence terms will display decay to equilibrium. As these are coupled to the diagonal elements of the reduced density matrix, these also relax, giving rise to the asymptotic limit of the microcanonical ensemble equilibrium state. The system is non-integrable and mixing leading to the initial conditions being totally forgotten and the establishment of the microcanonical equilibrium state. Therefore, to obtain a stable measurement record, the measured observable must be approximately conserved after measurement. This analysis emphasises why this is so important as if it is violated, the long time dynamics will not properly record a measurement. We know that the microcanonical ensemble is the attractor of the dynamics for the full isolated system as there are no other constants of the motion than the total energy. 

If we take the measuring apparatus to behave as a thermodynamic reservoir bath, we can approximate the equilibrium state of the subsystem through the canonical ensemble, in which case $\rho^{s}_{\text{eq}}\approx\frac{e^{-\beta H_s}}{Z_s}$, where $H_s$ is the Hamiltonian of the subsystem. However, this requires the measurement interaction to mimic the effects of a thermodynamic reservoir (weak coupling or through renormalization of the measurement apparatus effects on the system). Note that the general state  might not decompose further than the microcanonical ensemble restricted to the local observable whose total energy may be strongly coupled to the measuring apparatus. This is not possible for the case of a pure measurement Hamiltonian where the projections are conserved. As the intrinsic dynamics of the system does not preserve the state of the spins, we can have spin flips along with dephasing interactions. The full thermodynamic ensemble description is only relevant when the subsystem can induce transitions between the spin states, allowing the isolated system to genuinely exchange energy, thereby exploring the whole energy shell.

\end{fmffile}


\bibliography{main_article}

\end{document}